\documentclass[12pt,british]{elsarticle}
\usepackage[T1]{fontenc}
\usepackage[latin9]{inputenc}
\usepackage{geometry}
\usepackage[active]{srcltx}
\usepackage{units}
\usepackage{amstext}
\usepackage{amssymb}
\usepackage{graphicx}
\usepackage{esint}

\makeatletter
\usepackage{bm}
\def\vec#1{\bm{#1}}

\makeatother

\usepackage{babel}
\begin{document}

\begin{frontmatter}{}

\title{Alternative transient eddy-current\\
flowmetering methods for liquid metals}

\author{Richard Looney}

\ead{looneyr@uni.coventry.ac.uk }

\author{J\={a}nis Priede}

\ead{j.priede@coventry.ac.uk}

\address{Flow Measurement Research Centre, Coventry University, UK}
\begin{abstract}
\global\long\def\Rm{\mathit{Rm}}
We present a comprehensive numerical analysis of alternative transient
eddy-current flowmetering methods for liquid metals. This type of
flowmeter operates by tracking eddy-current markers excited by the
magnetic field pulses in the flow of a conducting liquid. Using a
simple mathematical model, where the fluid flow is replaced by a translating
cylinder, a number possible alternative measurement schemes are considered.
The velocity of the medium can be measured by tracking zero crossing
points and spatial or temporal extrema of the electromotive force
(emf) induced by transient eddy currents in the surrounding space.
Zero crossing points and spatial extrema of the emf travel synchronously
with the medium whereas temporal extrema experience an initial time
delay which depends on the conductivity and velocity of the medium.
Performance of transient eddy-current flowmetering depends crucially
on the symmetry of system. Eddy current asymmetry of a few per cent
makes the detection point drift with a velocity corresponding to a
magnetic Reynolds number $\Rm\sim0.1.$ With this level of asymmetry
transient eddy-current flowmetering can be reliably applicable only
to flows with $\Rm\gtrsim0.1.$ A more accurate symmetry adjustment
or calibration of flowmeters may be necessary at lower velocities.
\end{abstract}
\begin{keyword}
\noindent Electromagnetic flowmeter, liquid metal, eddy current 
\end{keyword}

\end{frontmatter}{}

\section{Introduction}

Accurate and reliable flowmetering of molten metals is required not
only in various metallurgical processes but also in the nuclear industry
where liquid metals are used for cooling of advanced reactors \citep{Schulenberg2010,Eckert2011,Poornapushpakala2014}.
The application of standard induction flowmeters to molten metals
is limited by their chemical aggressiveness which may cause corrosion
of electrodes and other contact problems. There is a variety of contactless
flowmeters which have been developed to avoid the problems with electrodes.
Induction flowmeters can be made contactless by using capacitately-coupled
electrodes \citep{Hussain1985,McHale1985}. However, most contactless
electromagnetic flowmeters for liquid metals employ various effects
related to eddy currents. For example, the flow rate can be determined
by measuring the force generated by eddy currents on a magnet placed
close to the flow of conducting liquid, as first suggested by \citet{Shercliff1987}
and recently pursued by the so-called Lorentz Force Velocimetry \citep{Thess2007a,Wegfrass2012}.
An alternative approach, which is virtually force-free and thus largely
independent of the conductivity of liquid metal \citep{Priede2009b},
is to determine the flow rate from the equilibrium rotation rate of
a freely rotating magnetic flywheel \citep{Shercliff1960,Bucenieks2005,Buchenau2014,Hvasta2018}
or just a single magnet \citep{Priede2011b}.

The standard eddy-current flowmeters operate by measuring the flow-induced
perturbation of the applied magnetic field \citep{Lehde1948,Cowley1965,Poornapushpakala2010a}.
The same principle underlies also the so-called flow tomography which
can reconstruct the basic features of the flow using the spatial distribution
of the induced magnetic field \citep{Stefani2004,Stefani2000}. The
application of this type of flowmeters becomes problematic when the
induced magnetic field is significantly weaker than the external field,
which is the case at low velocities. Although the background signal
produced by the transformer effect of the applied magnetic field can
be compensated by a proper arrangement of sending and receiving coils
\citep{Feng1975}, standard eddy-current flowmeters remain highly
susceptible to small geometrical imperfections and disturbances. The
sensitivity of eddy-current flowmeters to such geometrical disturbances
can significantly be reduced by measuring the phase shift induced
by the flow between two sensor coils instead of the usual amplitude
difference \citep{Priede2011c}. One of the remaining drawbacks of
the phase-shift flowmeter is the dependence of the signal not only
on the velocity but also on the conductivity of liquid metal. This
is a general problem which affects not only eddy-current but also
the Lorentz force flowmeters unless based on the elaborate time-of-flight
measurements \citep{Vire2010}. Recently, we showed that the sensitivity
of the phase-shift flowmeter to the variations of conductivity of
liquid metal can be significantly reduced by rescaling the flow-induced
phase shift between the receiving coils with the phase shift between
the sending and one of the receiving coils \citep{Looney2017}.

Another electromagnetic flowmeter, which is largely insensitive to
the conductivity of medium, is the pulsed field flowmeter \citep{Zheigur1965,Tarabad1983}.
This type of device, which has been recently developed under the name
of transient eddy-current flowmeter \citep{Forbriger2015} and claimed
to be calibration free \citep{Krauter2017a}, operates by exciting
and then tracking transient eddy current markers as they are carried
along by a moving conductor.

In this paper, we present a comprehensive numerical analysis of alternative
designs of transient eddy flowmeters which differ by the feature of
eddy current distribution tracked. The velocity of the medium can
be determined by tracking either zero crossing points or extrema of
the induced emf. There are two types of extrema \textendash{} spatial
and temporal, which can be tracked. We point out that the transient
eddy-current flowmetering relies essentially on the symmetry of the
system. 

The paper is organized as follows. In the next section, we introduce
a mathematical model of a transient eddy-current flowmeter where the
liquid flow is substituted by an infinite cylinder that translates
along its axis. The basics of the method are discussed in Sec. \ref{sec:eigm}
where the temporal evolution of axially mono-harmonic eddy current
eigenmodes are considered. In Section \ref{sec:Num} we present numerical
results for axially mono-harmonic eddy current distributions as well
as more realistic distributions generated circular current loops.
The paper is concluded by a summary and discussion of results in Section
\ref{sec:Sum}.

\section{\label{sec:Math}Mathematical Model}

\begin{figure}
\begin{centering}
\includegraphics[bb=100bp 80bp 370bp 290bp,clip,width=0.5\paperwidth]{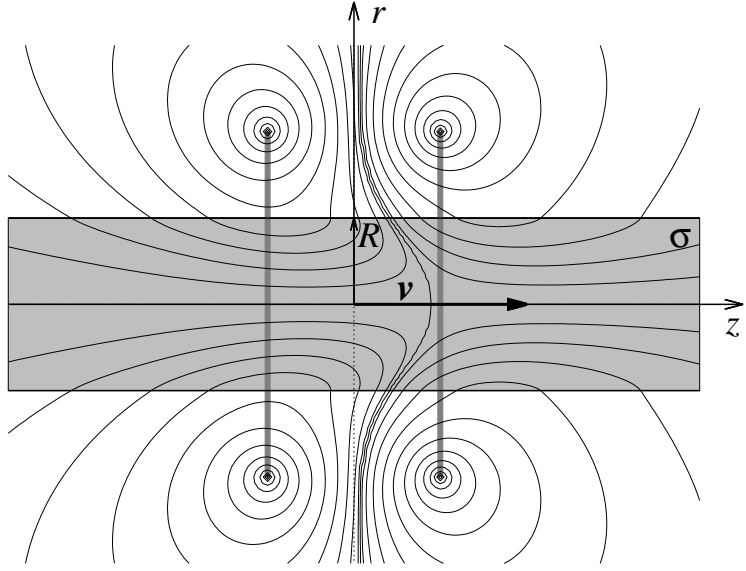}
\par\end{centering}
\caption{\label{cap:sketch}Sketch of set-up showing a cylinder of radius $R$
and electrical conductivity $\sigma$ translating at velocity $\vec{v}$
parallel to its axis in the magnetic field generated by two anti-symmetric
current loops. }
\end{figure}

Consider a solid infinitely long cylinder of radius $R$ and electrical
conductivity $\sigma$ translating at a velocity $\vec{v}=\vec{e}_{z}v$
parallel to its axis in an external magnetic field $\vec{B}_{0},$
which is periodically switched on and off for the time intervals $\tau$
and $T-\tau,$ respectively, where $T$ is the period of one full
cycle. The induced electric field is governed by the Maxwell-Faraday
equation 
\[
\vec{E}=-\vec{\nabla}\varphi-\partial_{t}\vec{A},
\]
where $\varphi$ is the electric potential and $\vec{A}$ is the vector
potential, which defines the magnetic field as $\vec{B}=\vec{\nabla}\times\vec{A}$.
The eddy current density induced in a moving medium is given by Ohm's
law 
\begin{equation}
\vec{j}=\sigma(\vec{E}+\vec{v}\times\vec{B})=\sigma(-\vec{\nabla}\varphi-\partial_{t}\vec{A}+\vec{v}\times\vec{\nabla}\times\vec{A}).\label{eq:ohm}
\end{equation}
 In the following, we consider an axisymmetric magnetic field which
has only $r$ and $z$ components in the cylindrical system of coordinates
and, thus, can be described by a purely azimuthal vector potential
$\vec{A}=\vec{e}_{\phi}A$ as $\vec{B}=-r^{-1}\vec{e}_{\phi}\times\vec{\nabla}(rA)$
with $\varphi\equiv0.$ Note that $\vec{B}\cdot\vec{\nabla}(rA)\equiv0,$
which means that the isolines of $rA=\Phi$ represent the flux lines
of $\vec{B}.$ Applying Ampere's law $\vec{j}=\mu_{0}^{-1}\vec{\nabla}\times\vec{B}$
to Eq. (\ref{eq:ohm}) we obtain the following advection-diffusion
equation for $A(\vec{r})$

\begin{equation}
\mu_{0}\sigma(\partial_{t}A+v\partial_{z}A)=\partial_{r}(r^{-1}\partial_{r}(rA))+\partial_{z}^{2}A,\label{eq:A}
\end{equation}
where $\mu_{0}=\unit[4\pi\times10^{-7}]{H/m}$ is the vacuum permeability.
The continuity of $\vec{B}$ at the cylinder surface at $r=1$ requires
the continuity of $A$ and $\partial_{r}A.$ 

Subsequently, we change to dimensionless variables by using $R,$
$\tau_{m}=\mu_{0}\sigma R^{2}$ and $v_{m}=(\mu_{0}\sigma R)^{-1}$
as the length, time and velocity scales, respectively. Then the problem
is defined by the magnetic Reynolds number $\Rm=\mu_{0}\sigma vR,$
which represents a dimensionless velocity.

We first consider the evolution of the eddy currents induced by the
external magnetic field in the form of a single Fourier harmonic which
varies as

\[
A_{0}(r,z,t)=\hat{A}_{0}(r)\sin(kz)f(t),
\]
where $k$ is the wave number and
\[
f(t)=\left\{ \begin{array}{ll}
\eta(t)+\eta(\tau-t)-\eta(T-t), & 0\le t<T\\
f(t-T), & t>T
\end{array}\right.=\sum_{n=-N/2}^{N/2}f_{n}e^{i\omega_{n}t}
\]
is the time variation which is defined using the complementary error
function $\eta(t)\text{=erfc}(t/\delta)$ to allow for a finite transition
time $\delta$ between the ``on'' and ``off'' states. This transition
time, which we set to three sampling time intervals $T/N$, is necessary
to suppress the Gibbs phenomenon in the Fourier series representation
of $f(t).$ The Fourier coefficients $f_{n}$ for the modes with frequencies
$\omega_{n}=2\pi n/T$ are computed using the FFT with a typical number
of sampling points $N=1024.$ The solution can be represented in the
complex form as 
\[
A(r,z,t)=\sum_{n=-N/2}^{N/2}\Im\left[\hat{A}_{n}(r)f_{n}e^{i(\omega_{n}t+kx)}\right].
\]
The radial distribution of the magnetic field outside the cylinder
is given by the general solution of Eq. (\ref{eq:A}) with $\sigma=0$
\begin{equation}
\hat{A}_{n}(r)=CI_{1}(kr)+D_{n}^{o}K_{1}(kr),\label{eq:Ano}
\end{equation}
where $I_{\nu}(x)$ and $K_{\nu}(x)$ are the modified Bessel functions
of the first and second kind with order $\nu$ \citep{Abramowitz1964};
$C$ is an unknown constant which depends on the current distribution
generating the field and $D_{n}^{o}$ is an unknown constant associated
with the $n$-th harmonic of the induced magnetic field. Inside the
cylinder, the solution of Eq. (\ref{eq:A}) which is regular at the
axis $r=0$ is 
\begin{equation}
\hat{A}_{n}(r)=D_{n}^{i}I_{1}(\kappa r),\label{eq:Ani}
\end{equation}
 where $\kappa=\left(k^{2}+i(\omega_{n}+k\Rm)\right)^{1/2}.$ The
unknown constants $D_{n}^{o}$ and $D_{n}^{i}$ are determined by
the continuity conditions of $A$ as follows 
\begin{eqnarray*}
D_{n}^{o} & = & C_{n}(kI_{0}(k)I_{1}(\kappa)-\kappa I_{0}(\kappa)I_{1}(k)),\\
D_{n}^{i} & = & C_{n}k(K_{0}(k)I_{1}(k)+I_{0}(k)K_{1}(k)),
\end{eqnarray*}
where $C_{n}=C/(kK_{0}(k)I_{1}(\kappa)+\kappa I_{0}(\kappa)K_{1}(k)).$
Since the current amplitude is irrelevant in our analysis, we can
set $C=1.$

The solution for a single Fourier harmonic obtained above can be extended
to a more realistic external magnetic field generated by a thin circular
loop. The free-space distribution of the magnetic field, which is
generated by a single current loop with radius $r_{c}$ and axial
position $z_{c}$ carrying the dimensionless current $j_{c}$, is
governed by 
\begin{equation}
\vec{\nabla}^{2}A_{0}=-j_{c}\delta(\vec{r}-r_{c}\vec{e}_{r}-z_{c}\vec{e}_{z}),\label{eq:A0}
\end{equation}
where $\delta(\vec{r})$ is the Dirac delta function and $\vec{r}$
is the radius vector. This problem can easily be solved using the
Fourier transform $\hat{A}(r)=\int_{-\infty}^{\infty}A(r,z)e^{ikz}\,dz,$
which converts Eq. (\ref{eq:A0}) into
\begin{equation}
(r^{-1}(r\hat{A}_{0})')'-k^{2}\hat{A_{0}}=-\hat{j}_{c}\delta(r-r_{c}),\label{eq:Ah0}
\end{equation}
where $\hat{j}_{c}=j_{c}e^{-ikz_{c}}.$ The solution, which is continuous
at $r=r_{c}$, regular at $r=0,$ and decays at $r\rightarrow\infty,$
can be written as 
\[
\hat{A}_{0}(r)=\left\{ \begin{array}{ll}
D_{c}I_{1}(kr)/I_{1}(kr_{c}), & r<r_{c};\\
D_{c}K_{1}(kr)/K_{1}(kr_{c}), & r>r_{c},
\end{array}\right.
\]
where $D_{c}=-\hat{j}_{c}/k\left(I_{0}(kr_{c})/I_{1}(kr_{c})+K_{0}(kr_{c})/K_{1}(kr_{c})\right)$
follows from the integration of Eq. (\ref{eq:Ah0}) over the singularity
at $r=r_{c}.$ Then the unknown constant defining the distribution
of the applied magnetic field in Eq. (\ref{eq:Ano}) can be written
as $C=\sum_{c}D_{c}/I_{1}(kr_{c})$ where the summation is over the
current loops generating the field. The vector potential in physical
space is obtained by the inverse Fourier transform which is computed
using the FFT with a typical number of sampling points $M=1024$ and
axial cut-off distance $z_{\max}=\pm51.2.$ 

\section{\label{sec:eigm}Eigenmode evolution}

The basics of transient eddy-current flowmetering are best revealed
by the evolution of separate eigenmodes, which can be sought in the
complex form as
\begin{equation}
A(r,z,t)=\hat{A}(r)e^{ikz-\gamma t},\label{eq:eigA}
\end{equation}
where $k$ is a given real wave number and $\gamma$ is an unknown
complex decay rate. The latter has to be determined together with
the amplitude distribution $\hat{A}(r)$ by solving the eigenvalue
problem posed by Eq. (\ref{eq:A}). In the absence of external magnetic
field, the solution outside the cylinder (\ref{eq:Ano}) reduces to
\begin{equation}
\hat{A}(r)=D^{o}K_{1}(kr),\label{eq:Ao}
\end{equation}
where $D^{o}$ is an unknown constant. Inside the cylinder, the general
solution of Eq. (\ref{eq:A}) can be written as 
\[
\hat{A}(r)=D^{i}J_{1}(\alpha r),
\]
where $D^{i}$ is another unknown constant, $J_{\nu}(x)$ is the Bessel
function of the first kind and order $\nu,$ and $\alpha=\left(\gamma-k^{2}-ik\Rm\right)^{1/2}.$
The continuity of $A$ and its first derivative leads to the following
characteristic equation
\begin{equation}
\alpha J_{0}(\alpha)/K_{0}(k)+kJ_{1}(\alpha)/K_{1}(k)=0,\label{eq:char}
\end{equation}
which has real roots $\alpha$ that define the complex associated
decay rates 
\begin{equation}
\gamma=\alpha^{2}+k^{2}+ik\Rm.\label{eq:gamma}
\end{equation}
The most important result that follows from this expression is the
constant phase speed $\Im[\gamma]/k=\Rm$ at which all eddy current
patterns travel regardless of their wave number. Note that the corresponding
physical velocity $\Rm v_{m}=v$ is that of the medium. This means
that the velocity of the medium can be determined by measuring the
phase velocity at which an eddy current pattern is advected. This
is the main idea behind the transient eddy-current flowmetering.

\begin{figure}
\centering{}\includegraphics[width=0.5\columnwidth]{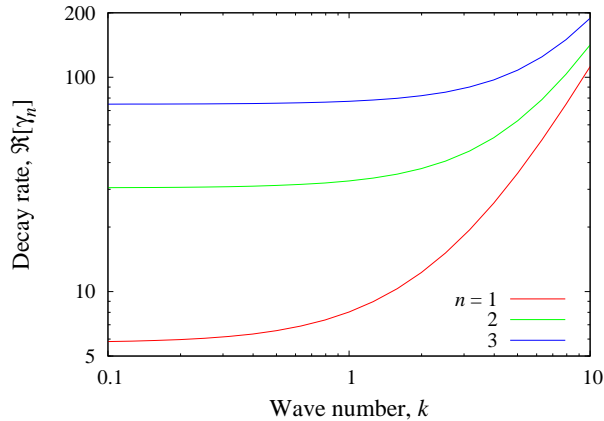}\caption{\label{fig:gamma}Three lowest eddy current decay rates versus the
wave number.}
\end{figure}

The second important result that follows from Eqs. (\ref{eq:char},\ref{eq:gamma})
concerns the decay rate $\Re[\gamma]$. As seen in figure \ref{fig:gamma},
where the decay rates of the first three dominant eigenmodes are plotted
against the wave number, the lowest decay rates occur in the long
wave limit $k\rightarrow0.$ In this limit, the characteristic equation
(\ref{eq:char}) reduces to $J_{0}(\kappa)=0$ and yields $\Re[\gamma_{1}]\approx5.78.$
It means that the eddy current amplitude drops by almost three orders
of magnitude over the the characteristic magnetic diffusion time $\tau_{m}=\mu_{0}\sigma R^{2}.$
The decay times of subsequent eigenmodes are significantly shorter.
It implies that the time interval over which a transient eddy current
pattern can be tracked is limited by a few magnetic diffusion time
scales $t_{m}.$ The respective dimensionless distance over which
the pattern is advected is, thus, limited by a few $\Rm.$ 

The decay of eddy currents makes the determination of their phase
speed more complicated than for a constant-amplitude wave. For the
latter, the phase velocity describes the motion of points with a fixed
amplitude. In a decaying wave, the only points whose amplitude remains
constant in time are those at which the oscillating quantity passes
through zero. Depending on the physical quantity whose zero crossing
is tracked, several alternatives of transient eddy-current flowmetering
are possible. Mathematically, the alternative quantities are related
with temporal or spatial derivatives of eddy current distribution.
For example, the emf induced by a decaying eddy current, which gives
rise to voltage in the pick-up coils, is related with the time derivative
of the associated magnetic flux. In our case, the latter is defined
by $\mathcal{E}=-\partial_{t}\Phi.$ Instead of zero crossings one
can also track extrema of $\mathcal{E}$, either in space or time,
which are defined mathematically by the zero crossings of $\partial_{z}\mathcal{E}$
and $\partial_{t}\mathcal{E},$ respectively. 

\section{\label{sec:Num}Results}

\subsection{Mono-harmonic eddy current distribution}

\begin{figure}
\centering{}\includegraphics[width=0.5\columnwidth]{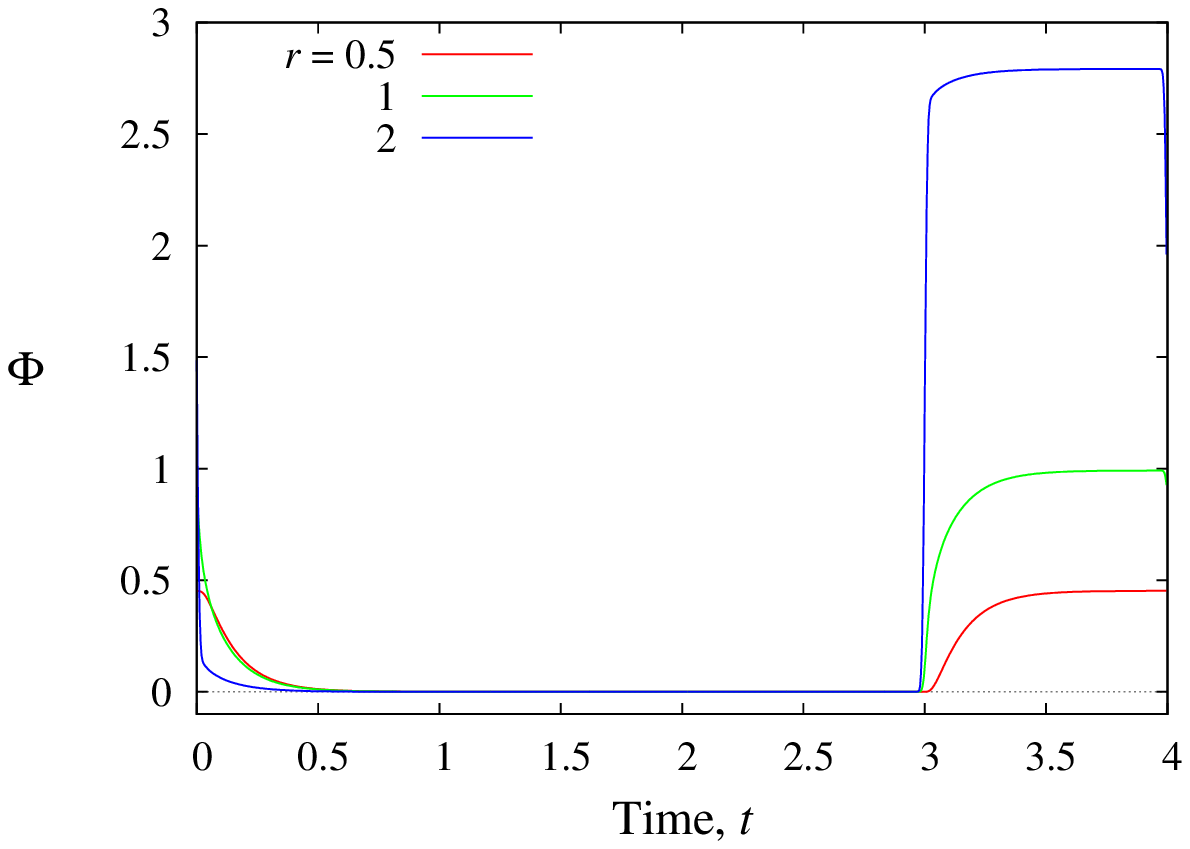}\put(-30,30){(\textit{a})}\includegraphics[width=0.5\columnwidth]{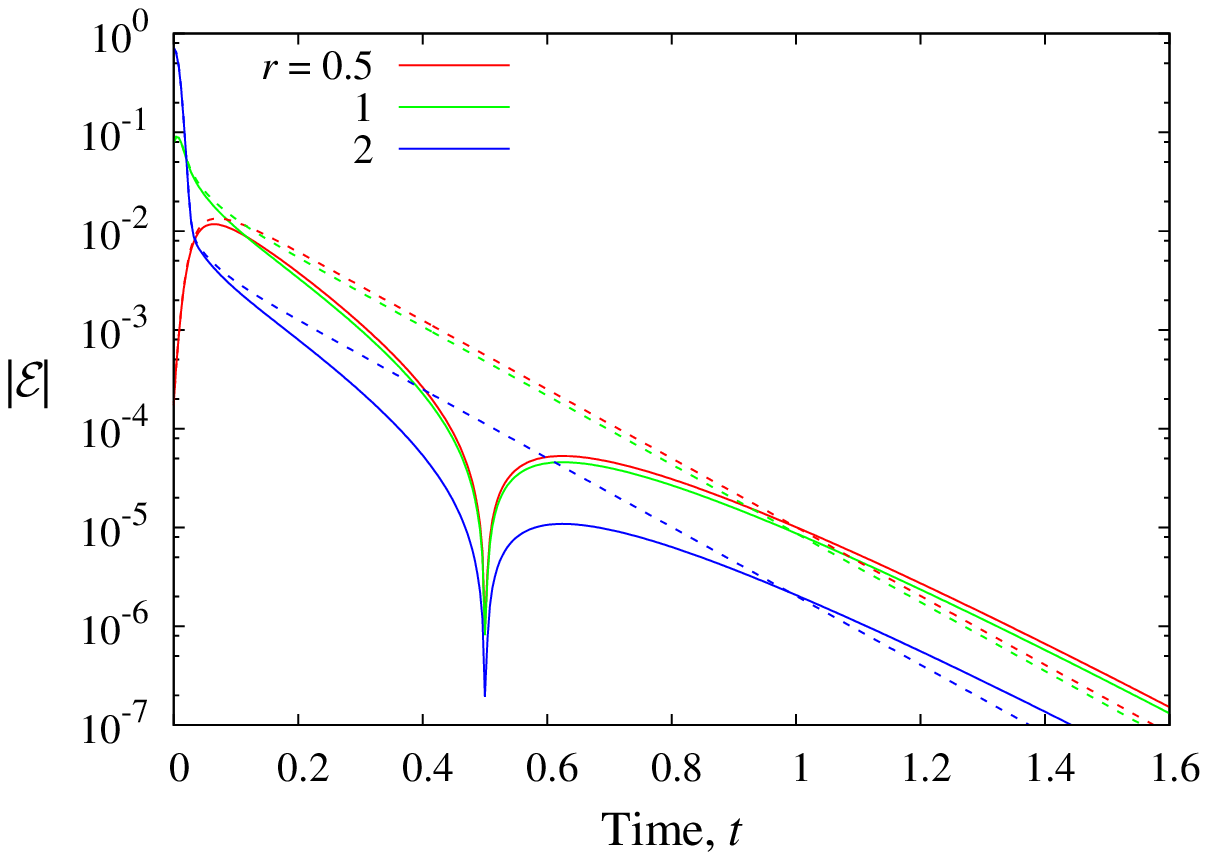}\put(-30,30){(\textit{b})}\caption{\label{fig:k1_t}Variation of the magnetic flux $\Phi=rA$ over one
time period at $z=0.5$ and $r=0.5,\,1,\,2$ for the Fourier harmonic
with the wave number $k=1$ when the cylinder is at rest $(\protect\Rm=0)$
(a); variation of the emf magnitude $|\partial_{t}\Phi|$ with time
at the same points for $\protect\Rm=0$ and $\protect\Rm=1$ (b).}
\end{figure}

Let us start with an external magnetic field which is periodically
switched off and on for the dimensionless time intervals $\tau=3$
and $T-\tau=1,$ respectively. According to the previous eigenvalue
analysis, these time intervals are sufficiently long for the eddy
currents to develop. This is confirmed by the temporal variation of
the magnetic flux $\Phi=rA$ shown in Fig. \ref{fig:k1_t}(a) for
the wave number $k=1$ at $z=0.5$ and three different radii when
the cylinder is at rest $(\Rm=0).$ The respective variation of the
emf magnitude is plotted in Fig. \ref{fig:k1_t}(b). When the cylinder
is at rest $(\Rm=0),$ the emf is seen to decrease exponentially with
time as predicted by the previous eigenvalue analysis. When the cylinder
moves with velocity $\Rm=1$ the decrease of emf is accompanied by
a zero crossing, which for the observation point located at $z=0.5$
occurs at the time instant $t\approx0.5.$ This point is seen as the
cusp on the semi-logarithmic plot of $|\mathcal{E}|$ in Fig. \ref{fig:k1_t}(b).
Shortly after passing through zero, emf is seen to attain a local
extremum, which is defined mathematically by zero crossing of $\partial_{t}\mathcal{E}.$

\begin{figure}
\begin{centering}
\includegraphics[bb=130bp 85bp 305bp 280bp,clip,width=0.25\columnwidth]{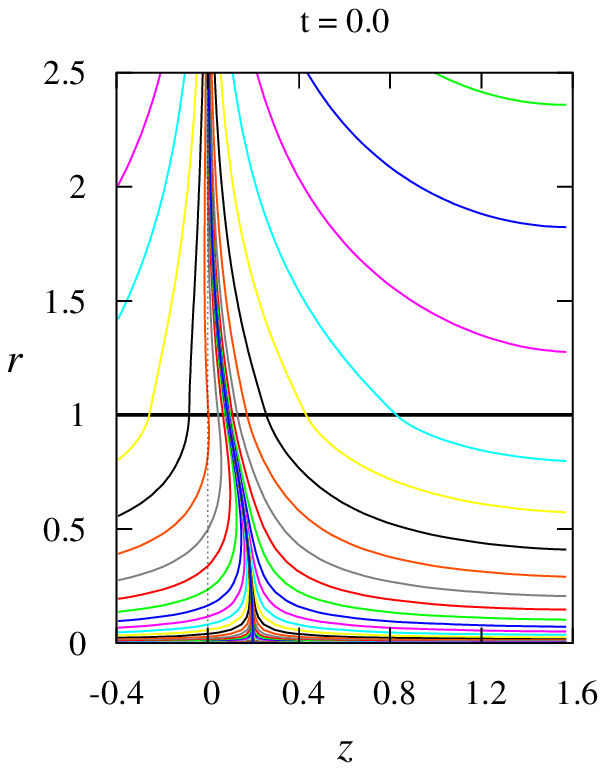}\includegraphics[bb=130bp 85bp 305bp 280bp,clip,width=0.25\columnwidth]{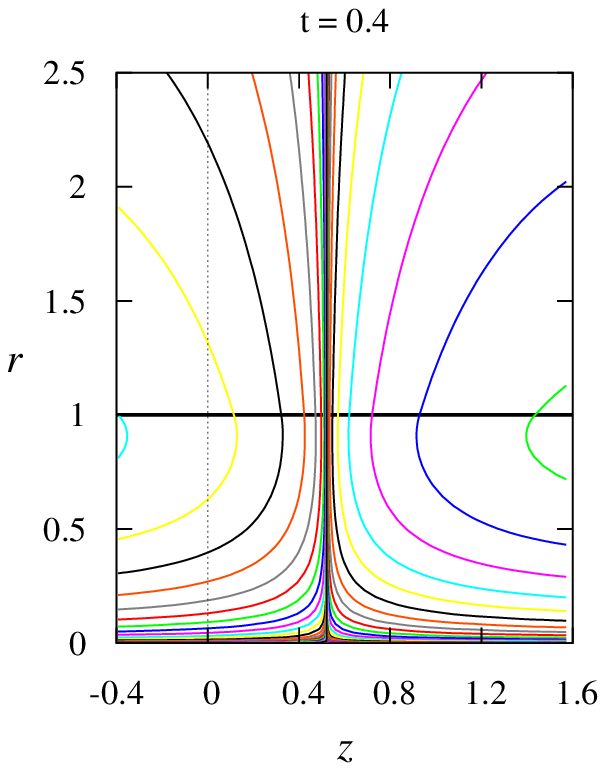}\includegraphics[bb=130bp 85bp 305bp 280bp,clip,width=0.25\columnwidth]{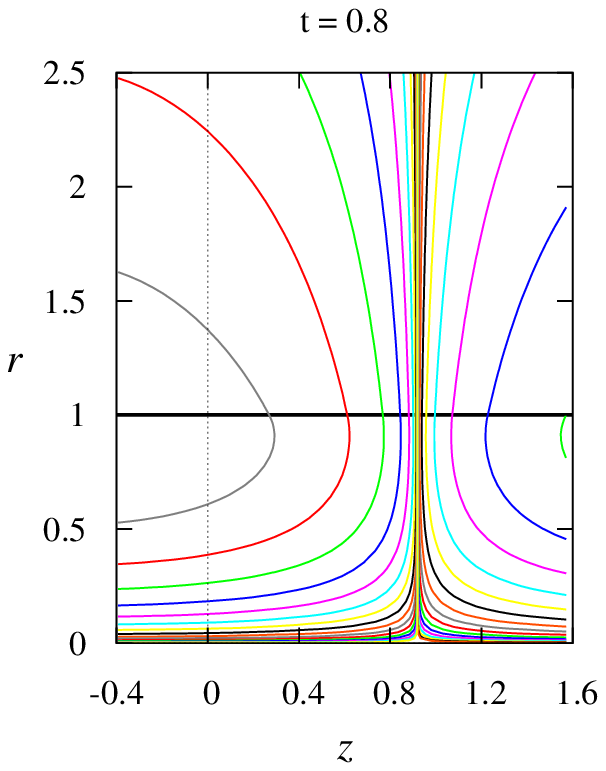}\includegraphics[bb=130bp 85bp 305bp 280bp,clip,width=0.25\columnwidth]{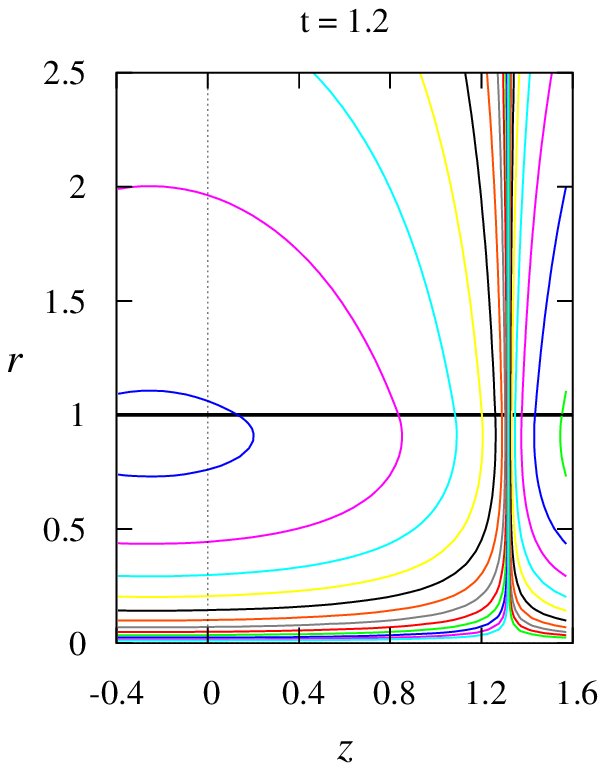}
\par\end{centering}
\begin{centering}
\includegraphics[bb=130bp 85bp 305bp 265bp,clip,width=0.25\columnwidth]{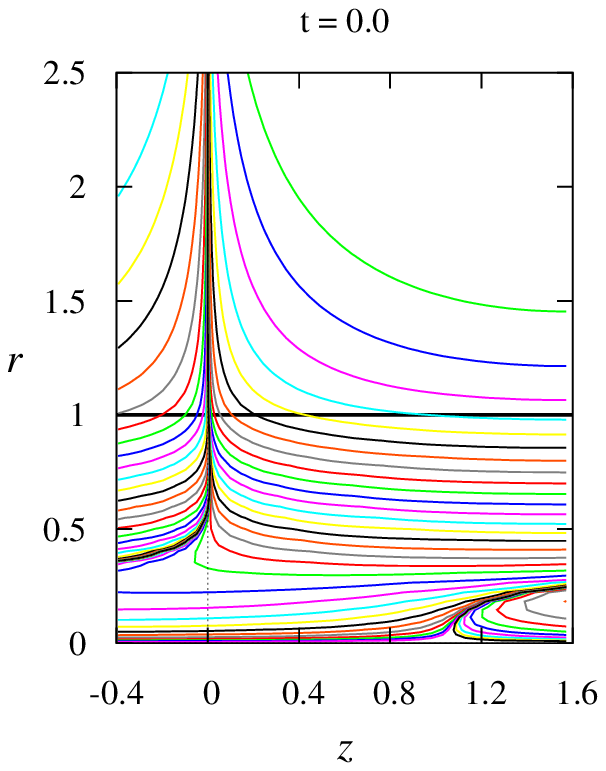}\includegraphics[bb=130bp 85bp 305bp 265bp,clip,width=0.25\columnwidth]{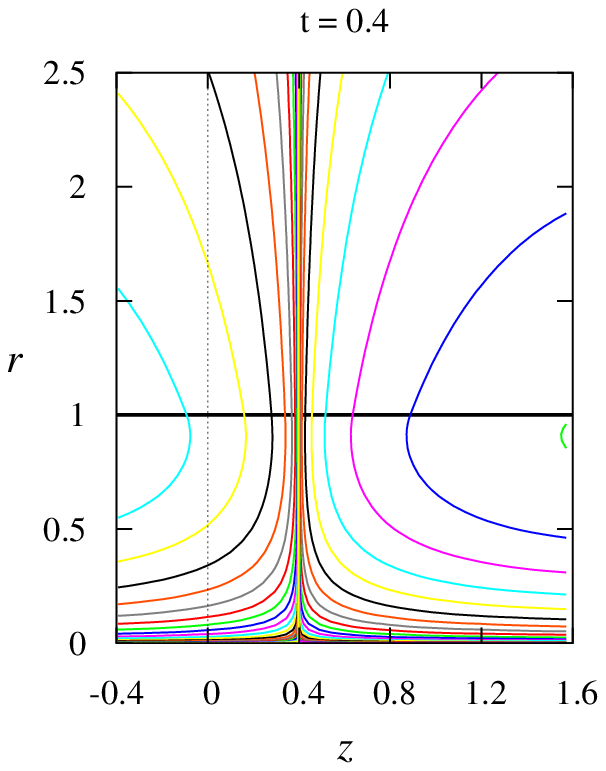}\includegraphics[bb=130bp 85bp 305bp 265bp,clip,width=0.25\columnwidth]{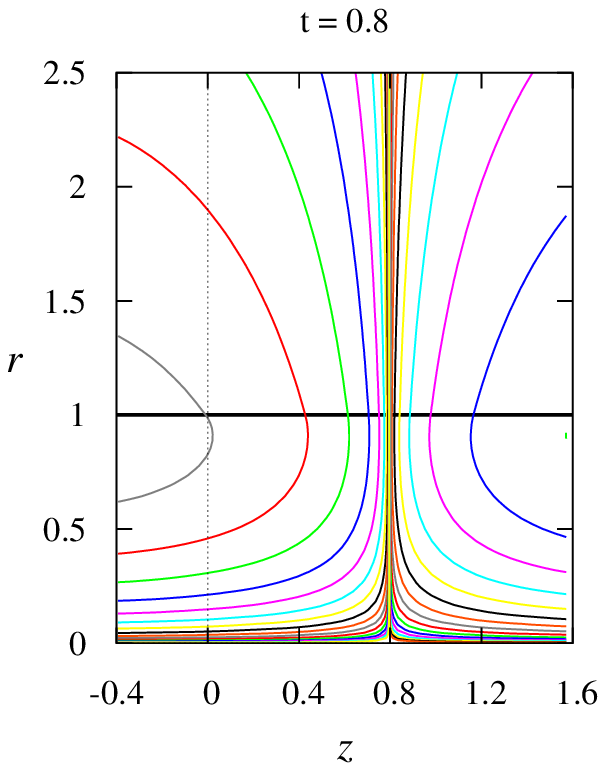}\includegraphics[bb=130bp 85bp 305bp 265bp,clip,width=0.25\columnwidth]{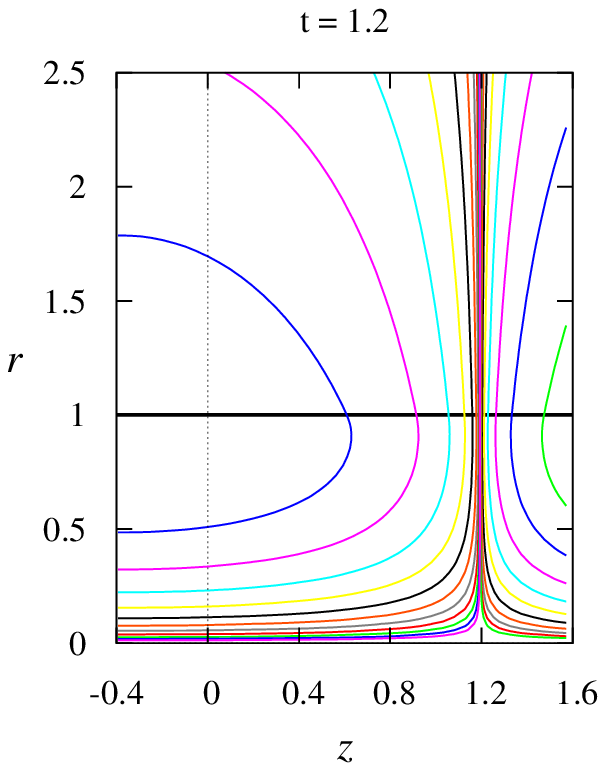}
\par\end{centering}
\begin{centering}
\includegraphics[bb=130bp 50bp 305bp 265bp,clip,width=0.25\columnwidth]{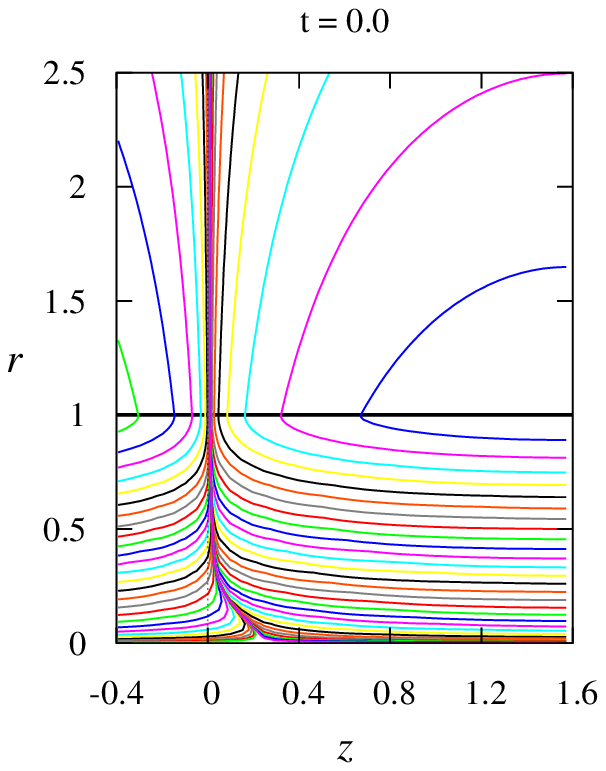}\includegraphics[bb=130bp 50bp 305bp 265bp,clip,width=0.25\columnwidth]{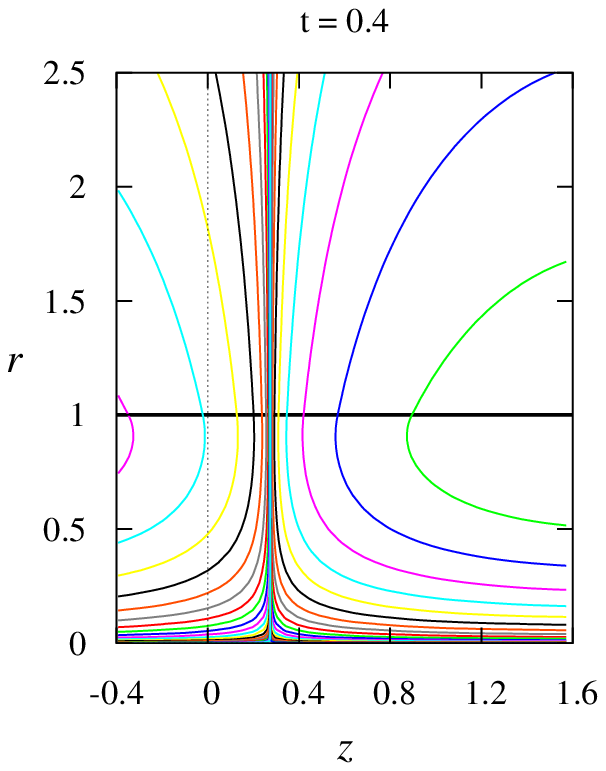}\includegraphics[bb=130bp 50bp 305bp 265bp,clip,width=0.25\columnwidth]{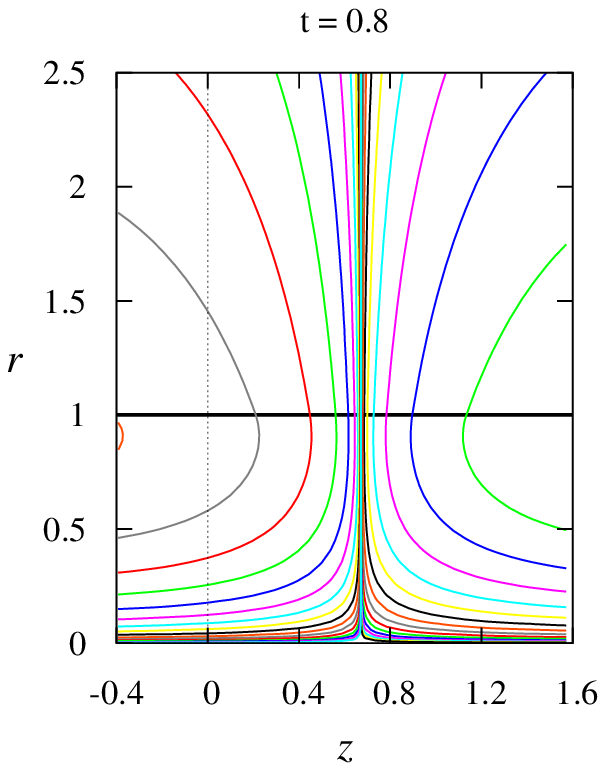}\includegraphics[bb=130bp 50bp 305bp 265bp,clip,width=0.25\columnwidth]{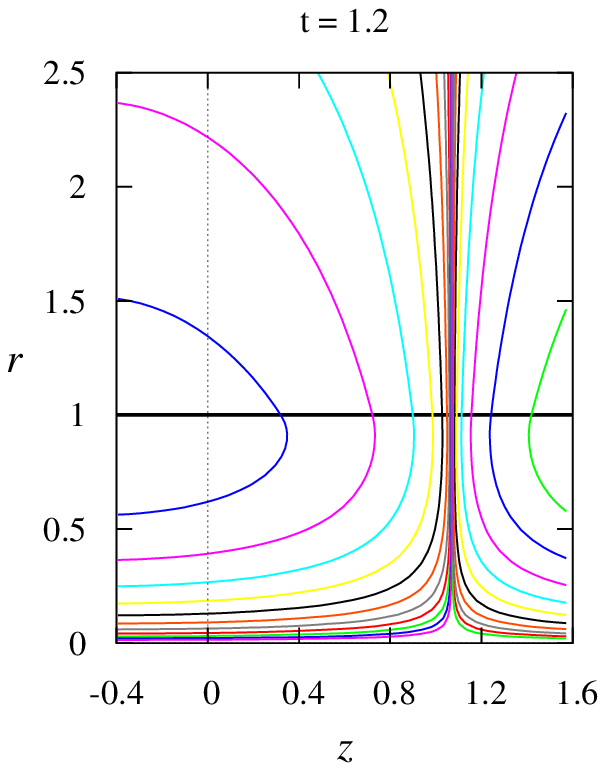}
\par\end{centering}
\centering{}\caption{\label{fig:u1k1}The magnetic flux lines $(\Phi=\text{const)}$ (top),
the isolines of emf $\mathcal{E}=-\partial_{t}\Phi$ (middle), and
of $\partial_{t}\mathcal{E}=-\partial_{tt}^{2}\Phi$ (bottom) for
$\protect\Rm=1$ at the time instants $t=0,\,0.4,\,0.8,\,1.2$ after
a mono-harmonic external magnetic field with the wave number $k=1$
has been switched off. Subsequent isolevels differ by a factor of
two and cluster around zero value.}
\end{figure}

Figure \ref{fig:u1k1} shows the evolution of the magnetic field pattern
and the associated emf with wave number $k=1$ after switching external
magnetic field off for the cylinder moving with velocity $\Rm=1.$
It may be seen that the zero crossing of the emf, which is marked
by the increased density of isolines in the middle row, closely follows
the medium by being located at $z=t\Rm.$ Thus, the velocity of medium
can be determined directly as $\Rm=z/t,$ where $z$ is the axial
distance of the observation point from the wave node and $t$ is the
time at which the emf passes through zero at that point after switching
the field off. The pattern of the magnetic flux lines, which is shown
at the top row of Fig. \ref{fig:u1k1}, may be seen to run slightly
ahead of that of the emf. This is obviously due to the effect of advection,
which tilts the magnetic flux lines in the direction of motion. On
the other hand, the time derivative, which is equivalent to the multiplication
of the dominating eigenmode (\ref{eq:eigA}) by $-\gamma,$ causes
a phase shift of the resulting distribution by $\arg(-\gamma).$ Thus,
the pattern of $\partial_{t}\mathcal{E}$, which is shown in the bottom
row of Fig. \ref{fig:u1k1}, lags slightly behind that of $\mathcal{E}.$
Note that the zero crossing of $\partial_{t}\mathcal{E},$ which like
that of $\mathcal{E}$ is marked by the increased density of the isolines,
indicates the location of temporal extremum of $\mathcal{E}.$ Location
of spatial extremum of $\mathcal{E}$ is defined by zero crossings
of $\partial_{z}\mathcal{E}.$ For a mono-harmonic eddy current, the
distribution of $\partial_{z}\mathcal{E}=ik\mathcal{E}$ is shifted
by a quarter wave length relative to that of $\mathcal{E}.$ Therefore
the spatial extrema of emf in a mono-harmonic wave move in exactly
the same way as zero crossings.

\begin{figure}
\centering{}\includegraphics[width=0.5\columnwidth]{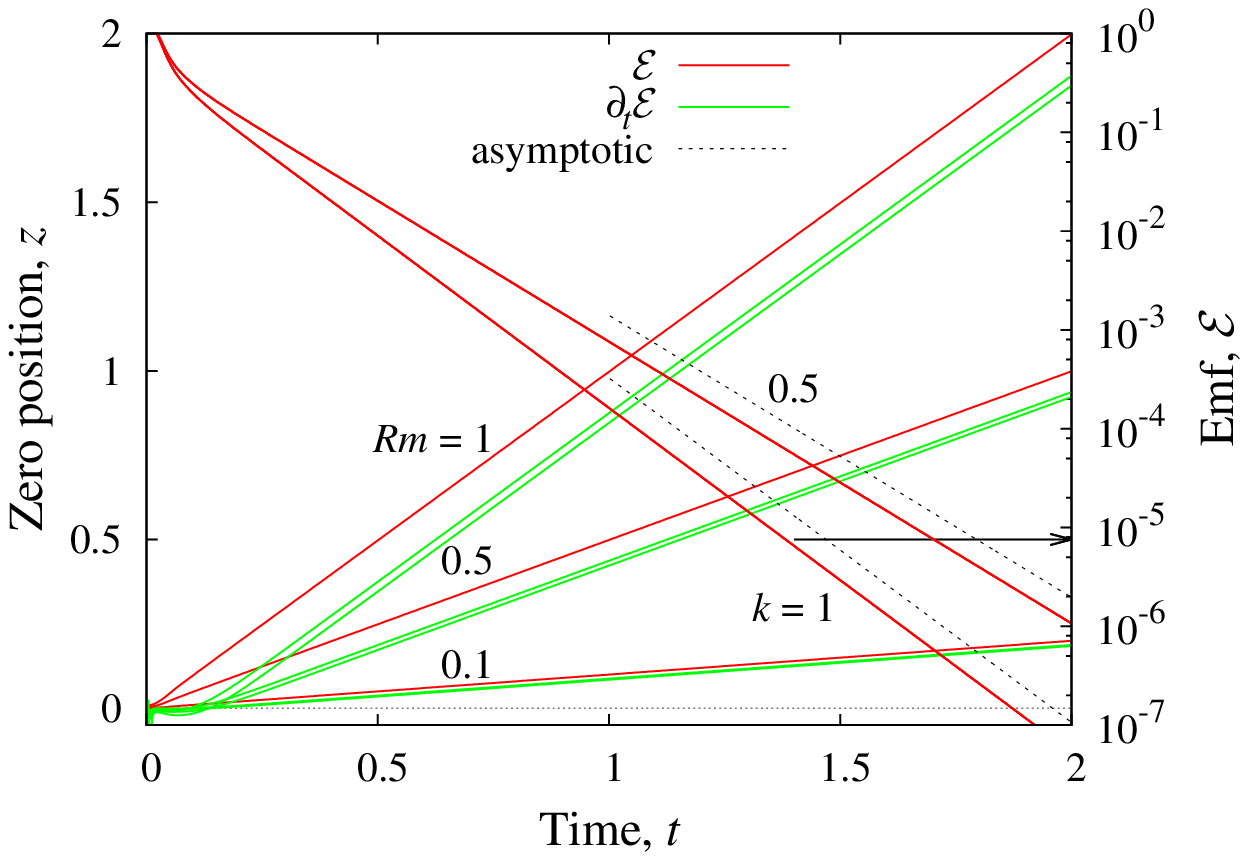}\put(-10,30){(\textit{a})}\includegraphics[width=0.5\columnwidth]{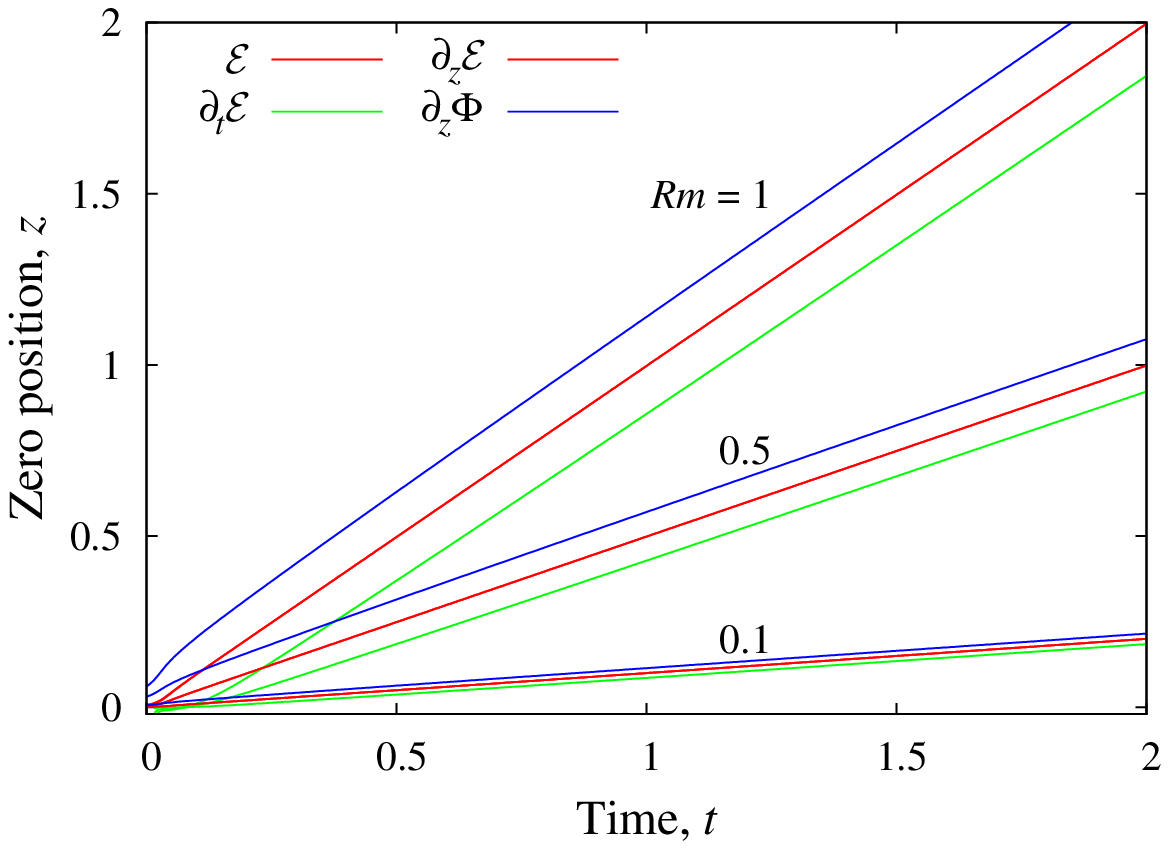}\put(-30,30){(\textit{b})}\caption{\label{fig:zero-t}(a) Axial zero crossing positions of $\mathcal{E}$
and $\partial_{t}\mathcal{E}$ as well as the relative magnitude of
$\mathcal{E}$ against time for mono-harmonic eddy current distributions
with $k=0.5,\,1$ and $\protect\Rm=0.1,\,0.5,\,1.$ (b) Axial zero
crossing positions of $\mathcal{E},\partial_{t}\mathcal{E}$ and $\partial_{z}\mathcal{E},\partial_{z}\Phi$
for the eddy current distributions generated respectively by two antisymmetric
and a single current loop with the radius $r=2$ and placed at $z=\pm1$
and $z=0.$ }
\end{figure}

As zero crossing outside the cylinder is seen in Fig. \ref{fig:u1k1}
to occur synchronously along the radius, in the following we focus
on the emf distribution along the surface $r=1.$ Figure \ref{fig:zero-t}(b)
shows zero crossing positions of both $\mathcal{E}$ and $\partial_{t}\mathcal{E}$
against time as well as the respective evolution of the emf amplitude
for two mono-harmonic eddy current distributions with wave numbers
$k=1$ and $k=0.5$ and three different velocities $\Rm=0.1,\,0.5,\,1.$
Firstly, the emf for both distributions may be seen to decay in a
good agreement with the analytically determined damping rates for
the respective wave numbers. Secondly, the zero crossing points of
$\mathcal{E}$ in both waves move in exactly the same way with velocity
$\Rm$ starting from the node $z=0$. Temporal extrema points, which
correspond to zero crossings of $\partial_{t}\mathcal{E},$ also move
at the same velocity as the medium but with a time delay which depends
on the wave number $k$ as well as on the velocity $\Rm$ itself.
It means that at least two measurement points are required to eliminate
this offset and, thus, to determine the velocity of the medium using
temporal extrema of emf.

\subsection{Eddy currents induced by circular loops}

\begin{figure}
\begin{centering}
\includegraphics[bb=55bp 100bp 380bp 240bp,clip,width=0.5\columnwidth]{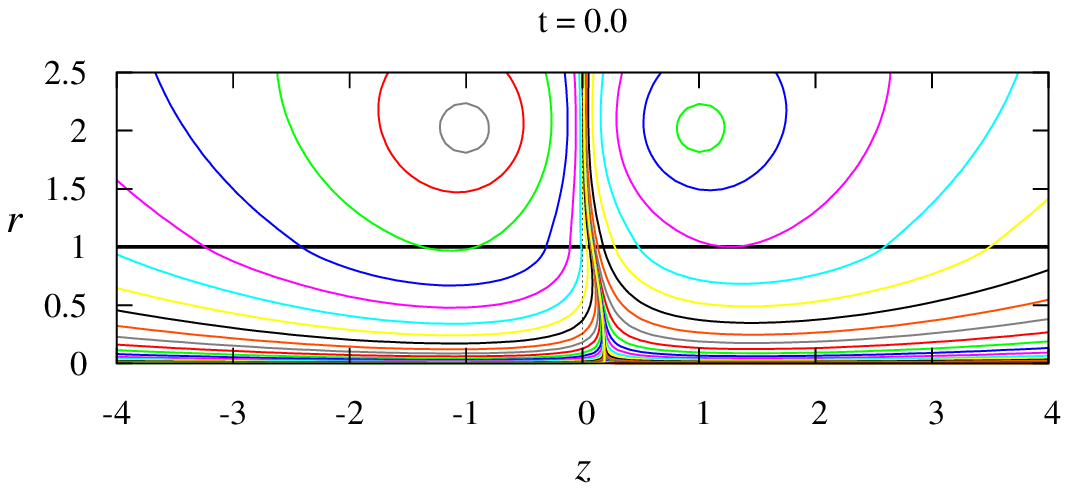}\includegraphics[bb=55bp 100bp 380bp 240bp,clip,width=0.5\columnwidth]{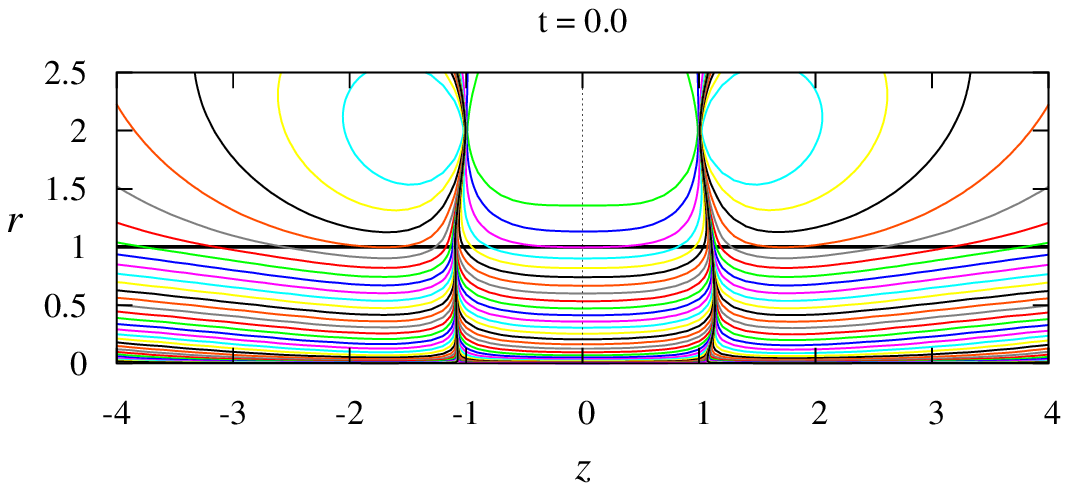}
\par\end{centering}
\begin{centering}
\includegraphics[bb=55bp 100bp 380bp 240bp,clip,width=0.5\columnwidth]{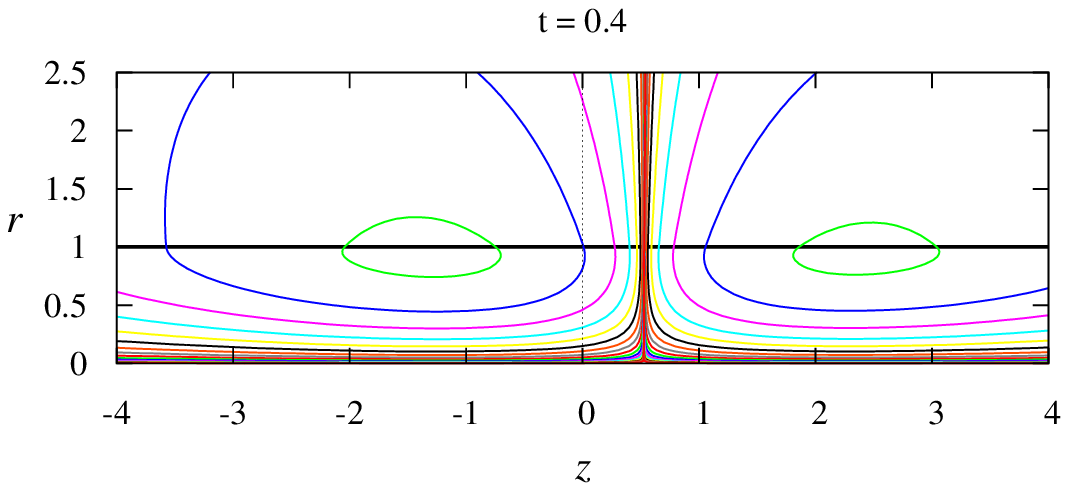}\includegraphics[bb=55bp 100bp 380bp 240bp,clip,width=0.5\columnwidth]{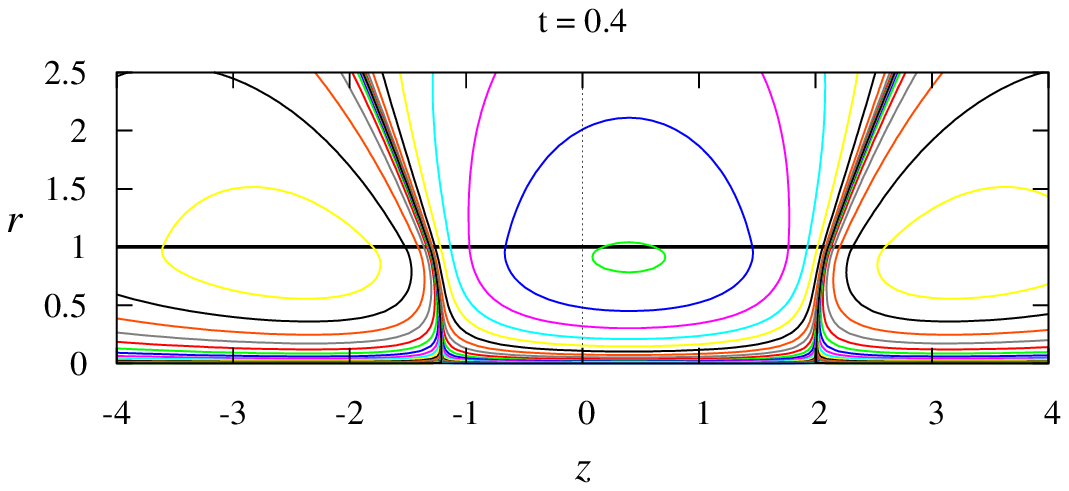}
\par\end{centering}
\begin{centering}
\includegraphics[bb=55bp 100bp 380bp 240bp,clip,width=0.5\columnwidth]{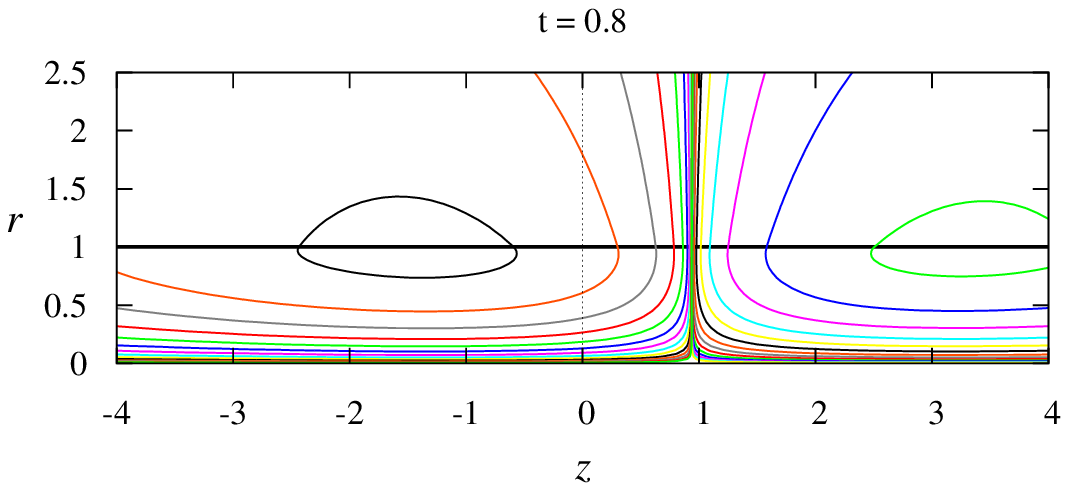}\includegraphics[bb=55bp 100bp 380bp 240bp,clip,width=0.5\columnwidth]{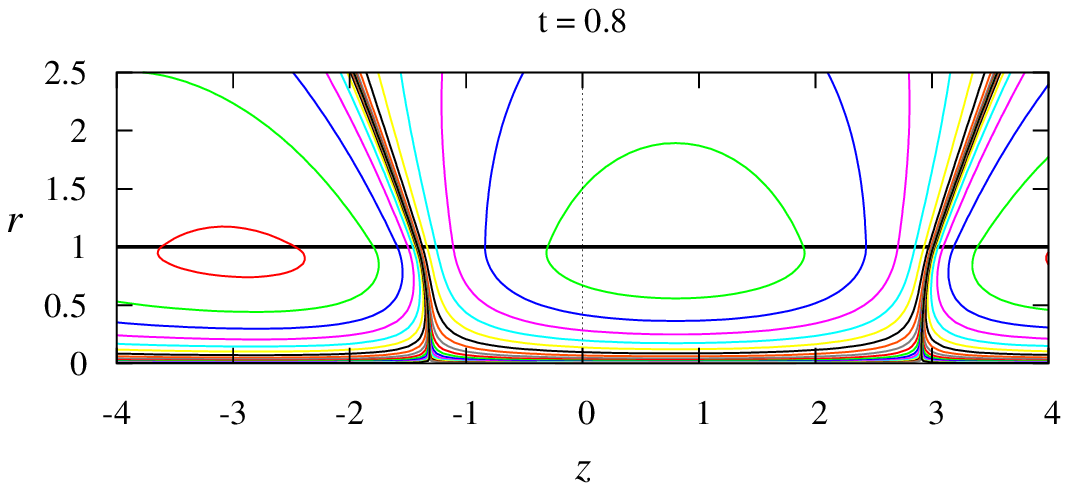}
\par\end{centering}
\begin{centering}
\includegraphics[bb=55bp 100bp 380bp 240bp,clip,width=0.5\columnwidth]{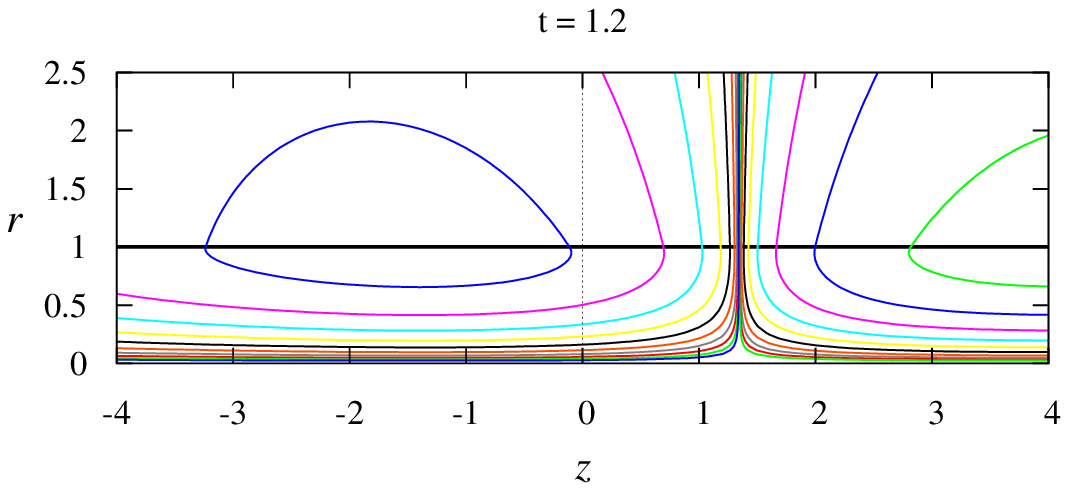}\includegraphics[bb=55bp 100bp 380bp 240bp,clip,width=0.5\columnwidth]{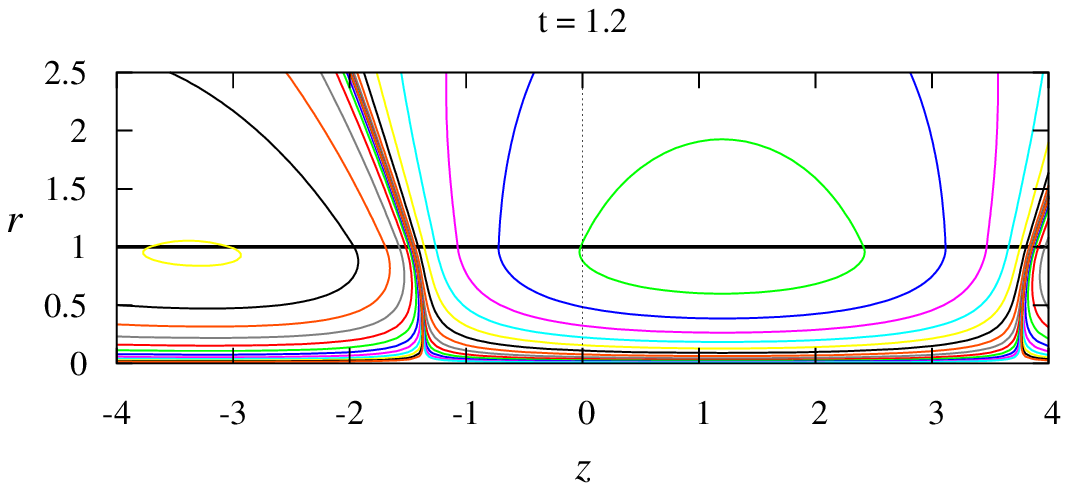}
\par\end{centering}
\centering{}\caption{\label{fig:u1r2z1x2}The magnetic flux lines $(\Phi=\text{const)}$
(left) and the isolines of $\partial_{z}\mathcal{E}=-\partial_{tz}^{2}\Phi$
(right) at the time instants $t=0,\,0.4,\,0.8,\,1.2$ after the external
magnetic field generated by a pair of antisymmetric current loops
with radius $r_{c}=2$ located at $z_{c}=\pm1$ has been switched
off. }
\end{figure}

Now let us consider the evolution of eddy currents generated by realistic
coils made of circular loops. We start with an antisymmetric coil
configuration consisting of two circular loops of radius $r_{c}=2$
which are placed at $z_{c}=\pm1$ and carry equal but opposite currents.
This configuration creates a zero crossing of emf at the symmetry
plane $z=0$ between the loops which is analogous to the wave node
of the mono-harmonic distribution considered in the previous section.
As a result, the advection of the field pattern by the moving medium,
which is shown for $\Rm=1$ in Fig. \ref{fig:u1r2z1x2}(left), is
similar to that of the mono-harmonic eddy current distribution in
Fig. \ref{fig:u1k1}(top). Also the zero crossing points of $\mathcal{E}$
and $\partial_{t}\mathcal{E}$ move in the same way as in the mono-harmonic
wave. But there is one substantial difference between the mono-harmonic
and anti-symmetric eddy-current distributions which concerns the motion
of spatial extrema of emf. There are two such extrema, which are seen
in Fig. \ref{fig:u1r2z1x2}(right) to be located at the current loops
where zero crossings of $\partial_{z}\mathcal{E}$ are marked by the
increased density of isolines. First of all, it is obvious that these
extrema do not move at the same velocity. Namely, the right (downstream)
extremum moves noticeably faster than the medium whereas the left
(upstream) one moves not only much slower but also in the opposite
direction. The main difference between the spatial extrema in the
previous mono-harmonic and the present two-loop eddy current distributions
is the absence of symmetry in the latter. It will be shown later that
symmetry is crucial to the transient eddy current flowmetering.

\begin{figure}
\begin{centering}
\includegraphics[bb=130bp 85bp 305bp 280bp,clip,width=0.25\columnwidth]{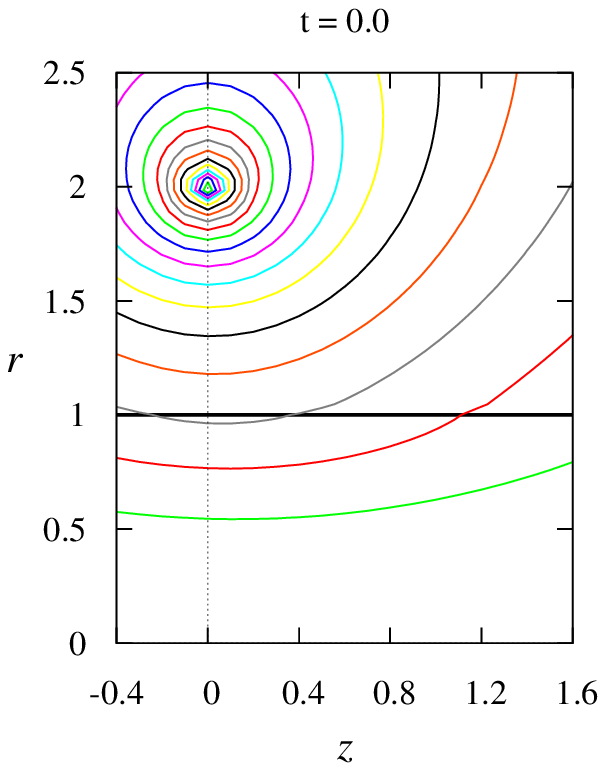}\includegraphics[bb=130bp 85bp 305bp 280bp,clip,width=0.25\columnwidth]{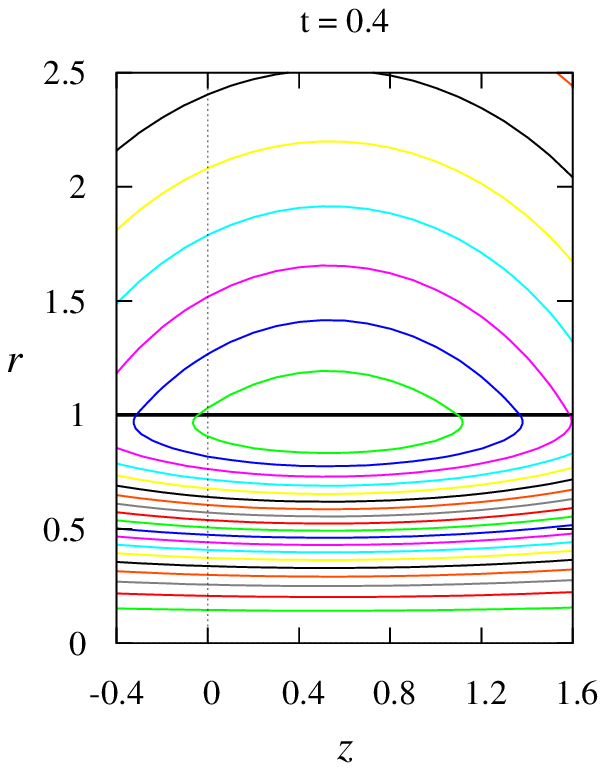}\includegraphics[bb=130bp 85bp 305bp 280bp,clip,width=0.25\columnwidth]{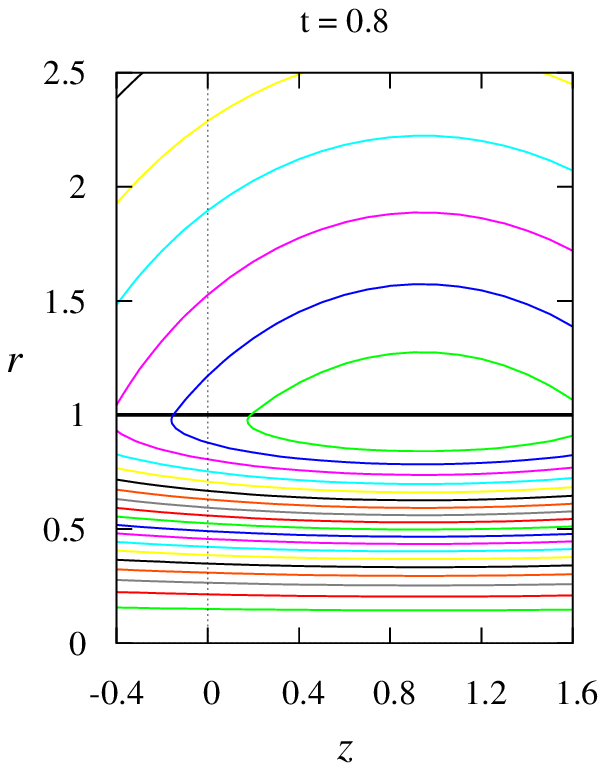}\includegraphics[bb=130bp 85bp 305bp 280bp,clip,width=0.25\columnwidth]{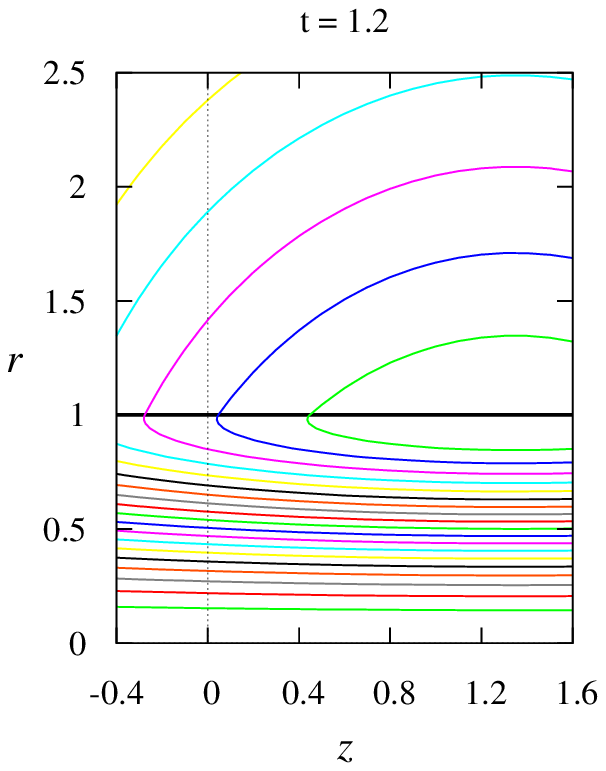}
\par\end{centering}
\begin{centering}
\includegraphics[bb=130bp 85bp 305bp 265bp,clip,width=0.25\columnwidth]{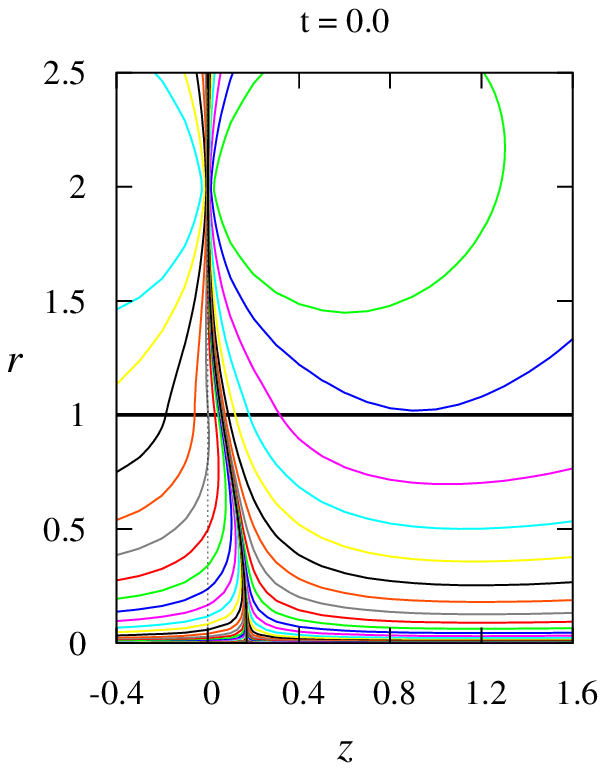}\includegraphics[bb=130bp 85bp 305bp 265bp,clip,width=0.25\columnwidth]{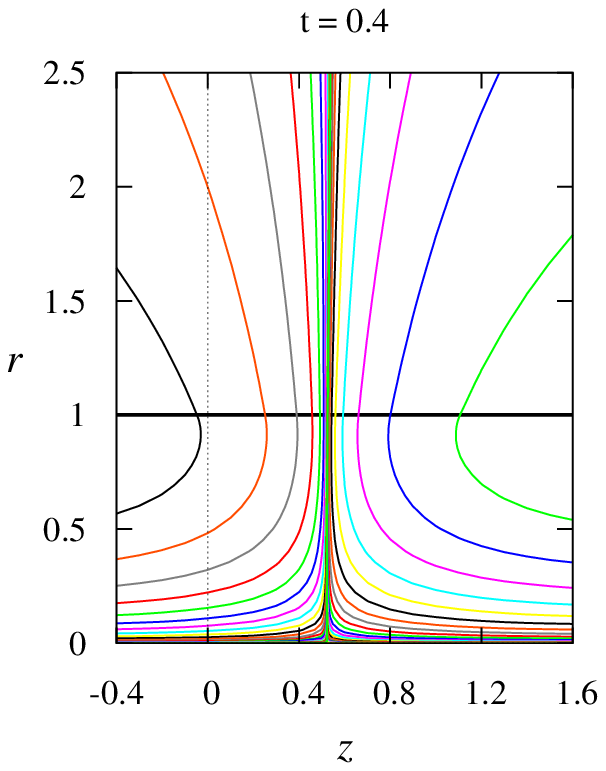}\includegraphics[bb=130bp 85bp 305bp 265bp,clip,width=0.25\columnwidth]{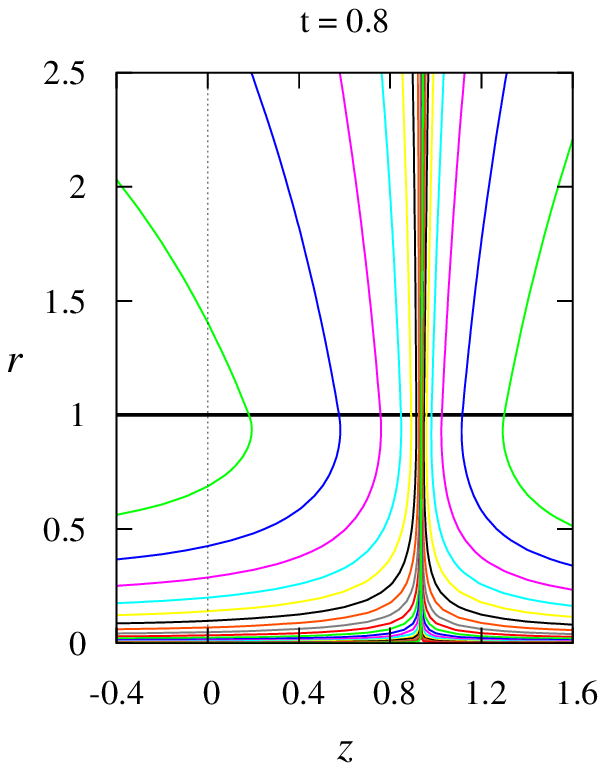}\includegraphics[bb=130bp 85bp 305bp 265bp,clip,width=0.25\columnwidth]{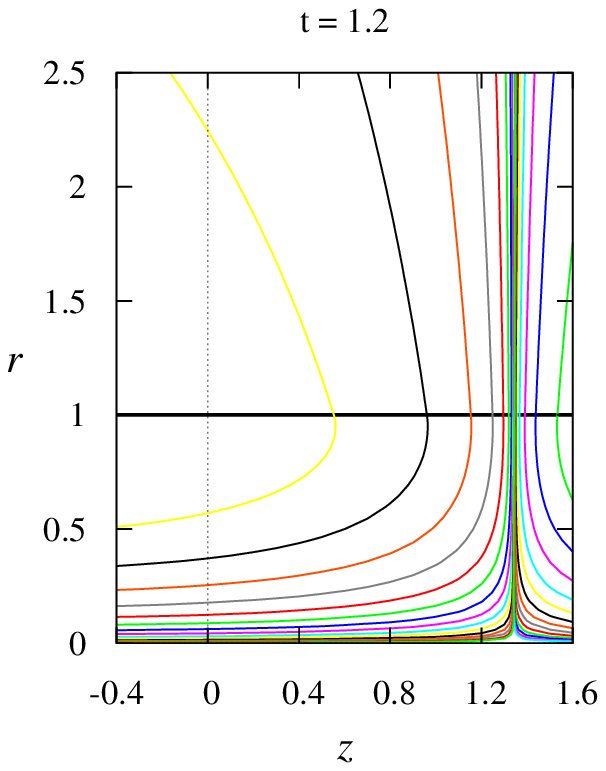}
\par\end{centering}
\begin{centering}
\includegraphics[bb=130bp 50bp 305bp 265bp,clip,width=0.25\columnwidth]{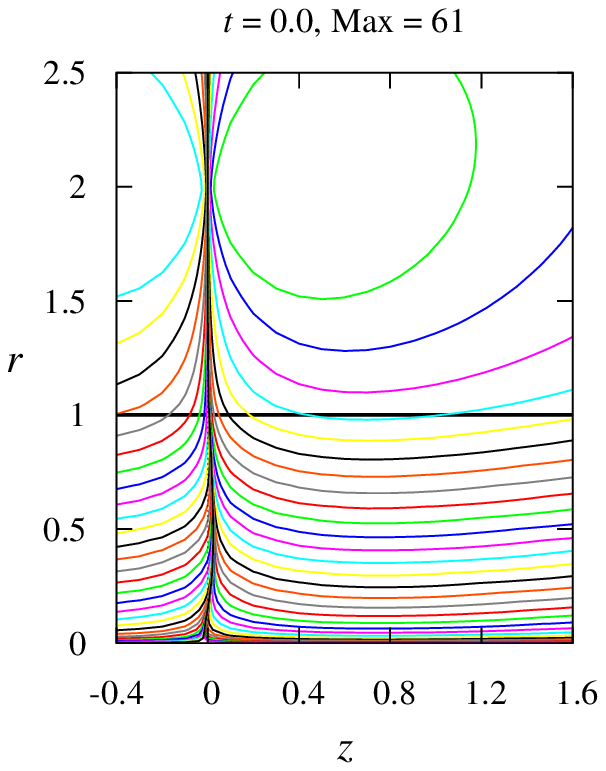}\includegraphics[bb=130bp 50bp 305bp 265bp,clip,width=0.25\columnwidth]{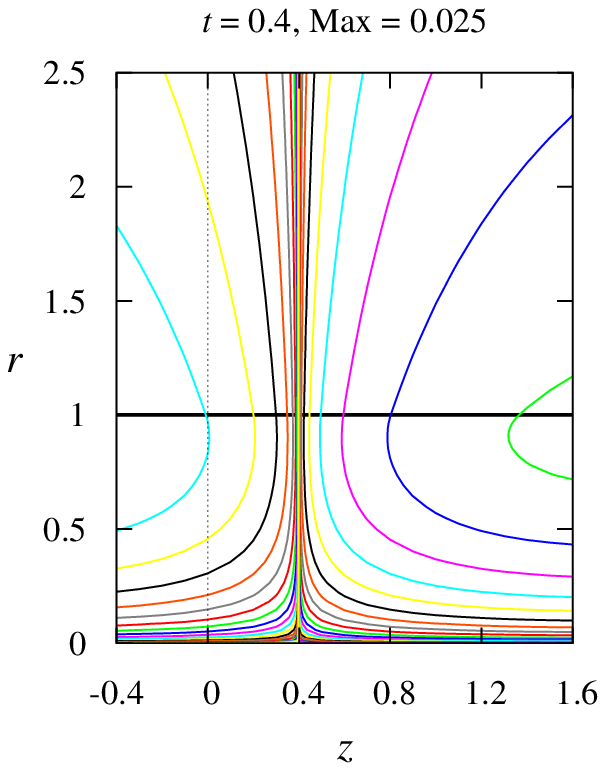}\includegraphics[bb=130bp 50bp 305bp 265bp,clip,width=0.25\columnwidth]{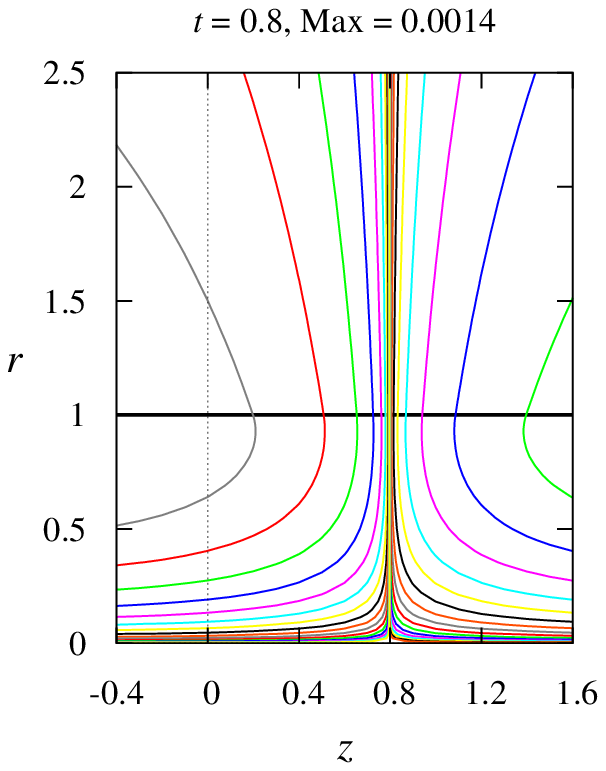}\includegraphics[bb=130bp 50bp 305bp 265bp,clip,width=0.25\columnwidth]{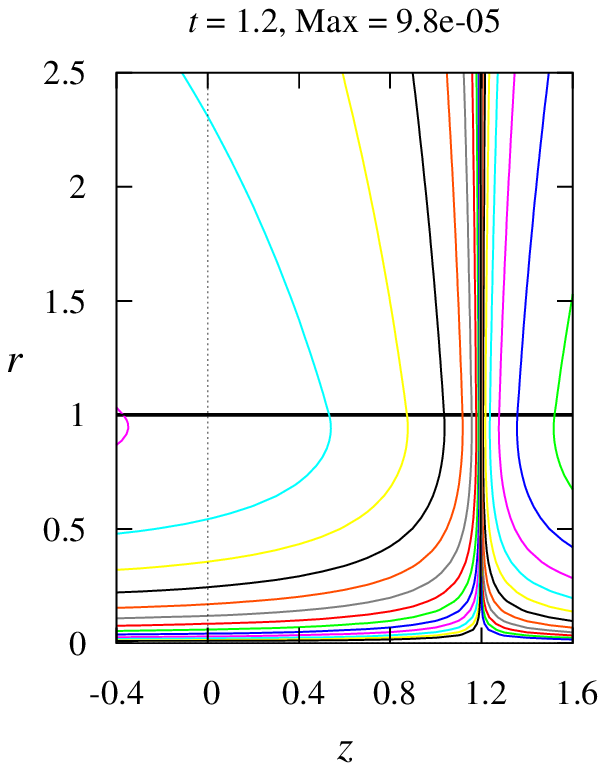}
\par\end{centering}
\centering{}\caption{\label{fig:u1r2z0x1}The magnetic flux lines $(\Phi=\text{const)}$
(top), the isolines of $\partial_{z}\Phi$ (middle) and $\partial_{z}\mathcal{E}=-\partial_{zt}^{2}\Phi$
(bottom) for $\protect\Rm=1$ at the time instants $t=0,\,0.4,\,0.8,\,1.2$
after the external magnetic field generated by a single current loop
located at $r_{c}=2$ and $z_{c}=0$ has been switched off. Levels
of subsequent isolines differ by a factor of two and the increased
density of isolines indicates zero value.}
\end{figure}

Eddy current distribution with a spatially symmetric emf extremum
but without zero crossing can be generated using a single current
loop \citep{Tarabad1983}. The evolution of the field pattern generated
by a current loop of radius $r_{c}=2$ placed at $z_{c}=0$ is shown
in Fig. \ref{fig:u1r2z0x1} for $\Rm=1.$ In this case, there are
neither zero crossings nor temporal extrema of emf but only axial
extrema of the magnetic flux and emf. These extrema, which are respectively
located at the zero crossings of $\partial_{z}\Phi=-rB_{r}$ and $\partial_{z}\mathcal{E},$
move synchronously with the medium. The axial extremum of emf is seen
to move without the time lag as the zero crossing in the anti-symmetric
set-up, whereas the flux extremum experiences a time lag similar to
that of the temporal emf maximum in the anti-symmetric set-up. Note
that the axial extremum of the magnetic flux can be detected as a
zero crossing of the radial flux component $B_{r}$ using, for example,
a Hall sensor. Also note that at least two sensor coils are required
to detect an axial maximum of emf, whereas one coil can be used to
detect zero crossing or temporal extremum of emf in the antisymmetric
set-up. The latter, however, requires two excitation coils.

\begin{figure}
\begin{centering}
\includegraphics[bb=130bp 50bp 305bp 280bp,clip,width=0.25\columnwidth]{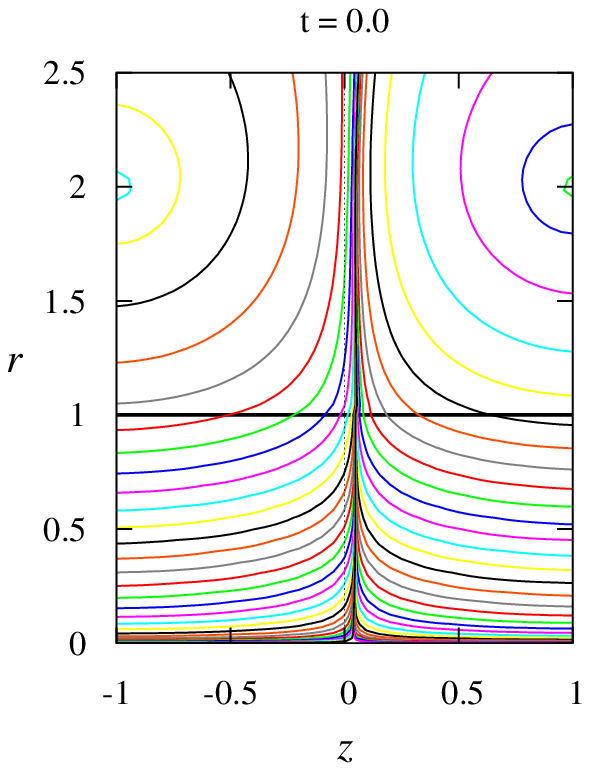}\includegraphics[bb=130bp 50bp 305bp 280bp,clip,width=0.25\columnwidth]{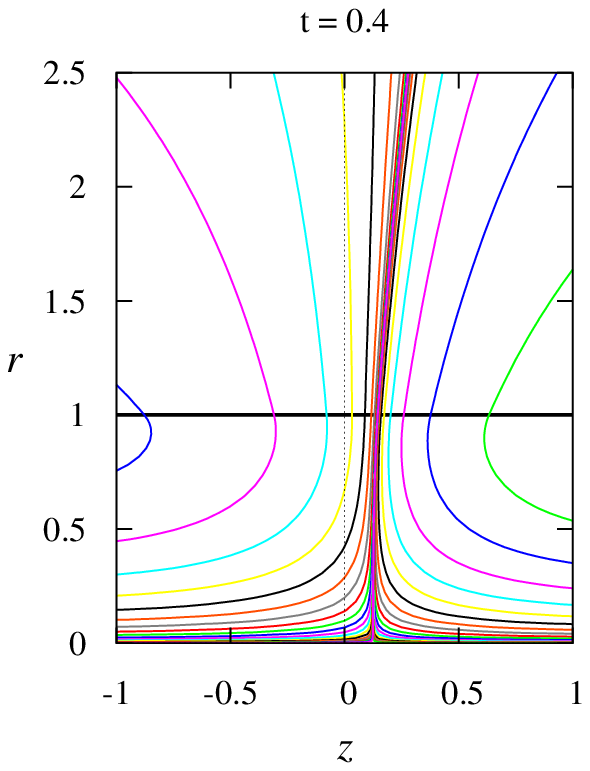}\includegraphics[bb=130bp 50bp 305bp 280bp,clip,width=0.25\columnwidth]{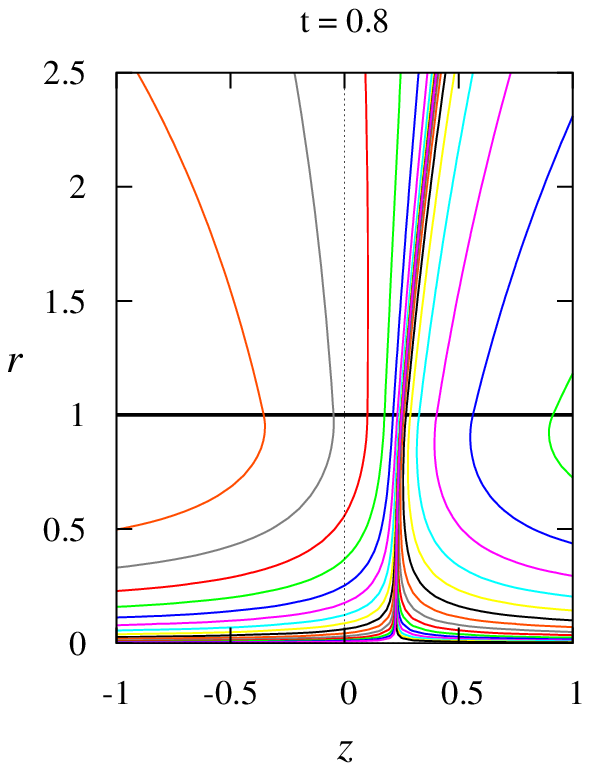}\includegraphics[bb=130bp 50bp 305bp 280bp,clip,width=0.25\columnwidth]{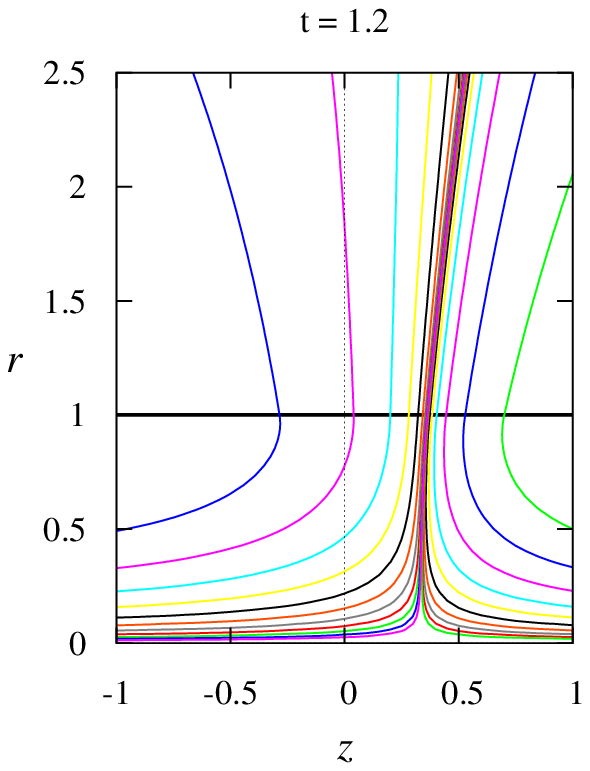}
\par\end{centering}
\centering{}\caption{\label{fig:u0r2z1x2+5}The emf isolines for $\protect\Rm=0$ at the
time instants $t=0,\,0.4,\,0.8,\,1.2$ after the external magnetic
field generated by a pair of opposite current loops located at $r_{c}=2$
and $z_{c}=\pm1$ with the current asymmetry of $S=5\%$ has been
switched off.}
\end{figure}

\begin{figure}
\centering{}\includegraphics[width=0.5\columnwidth]{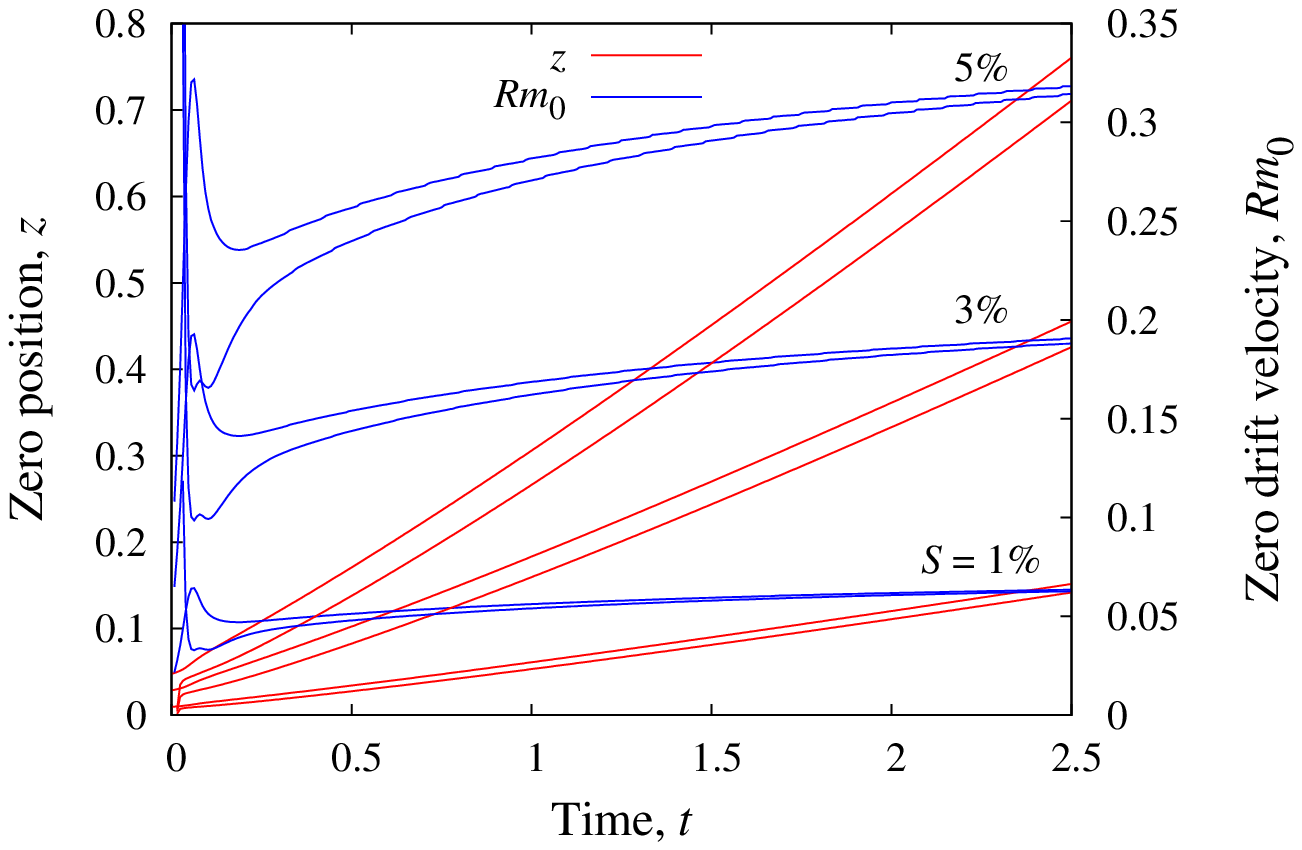}\put(-50,30){(\textit{a})}\includegraphics[width=0.5\columnwidth]{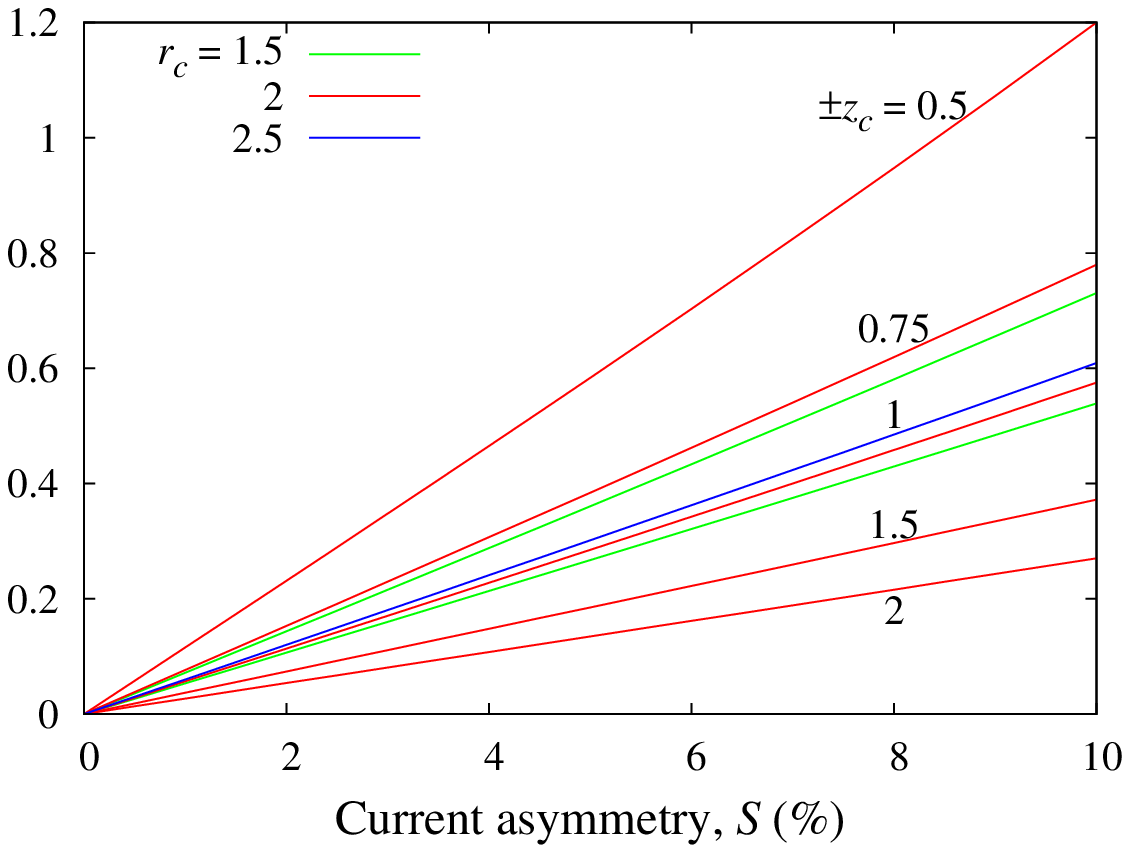}\put(-30,30){(\textit{b})}\caption{\label{fig:v0_r2z1x2}(a) Axial position of zero crossing and its
drift velocity along the surface of cylinder at rest $(\protect\Rm=0)$
against the time after the external magnetic field generated by a
pair of opposite current loops located at $r_{c}=2$ and $z_{c}=\pm1$
with the current asymmetry $S$ has been switched off. The upper and
lower curves correspond to the zero crossing of emf $($$\mathcal{E}=-\partial_{t}\Phi)$
and its temporal derivative $($$\partial_{t}\mathcal{E})$, respectively.
(b) Zero drift velocity depending on the current asymmetry $s$ in
two circular loops with radii $r_{c}=1.5,\,2,\,2.5$ placed at the
axial positions $\pm z_{c}=0.5,\,0.75,\,1,\,1.5,\,2.$ }
\end{figure}

Finally, let us examine the effect of a possible asymmetry in the
initial eddy current distribution generated by a two-coil set-up with
opposite but slightly different currents. To characterize this kind
of asymmetry we use the parameter $S=(j_{+}-j_{-})/(j_{+}+j_{-}),$
where $j_{+}$ and $j_{-}$ are the currents in the coils placed respectively
on the left and right from $z=0.$ Temporal evolution of eddy current
distribution with the initial asymmetry of $S=5\%$ generated by two
coils of radius $r_{c}=2$ placed $z_{c}=\pm1$ is shown in Fig. \ref{fig:u0r2z1x2+5}
for the medium at rest $(\Rm=0).$ Because for $S>0$ the current
in the coil on the left is higher than that on the right the initial
emf pattern at $t=0$ is slightly tilted to the right. In contrast
to the perfectly anti-symmetric distribution, where each Fourier mode
of emf crosses zero at $z_{c}=0$ independently of other harmonics,
in the asymmetric distribution, the zero crossing is a result of superposition
of different Fourier modes. Because different harmonics decay at different
rates depending on their wave number, the zero crossing line in the
asymmetric distribution is not stationary but drifts to the right,
as seen in Fig. \ref{fig:u0r2z1x2+5}. The direction of this drift
is reversed for negative $S.$ As seen in Fig. \ref{fig:v0_r2z1x2},
after a relatively short initial transient time, the drift velocity
slightly increases and then saturates at the level which rises with
the asymmetry $S$ and is nearly the same for zero crossing and temporal
extremum of emf. The drift velocity averaged over the time interval
from $t=0$ to $t=3$ is seen in Fig. \ref{fig:v0_r2z1x2} to increase
nearly linearly with $S.$ At the same time the drift velocity reduces
with the increase of axial separation between the coils whereas their
radius has a relatively weak effect.

\section{\label{sec:Sum}Summary and conclusions}

We have carried out a comprehensive numerical analysis of a transient
eddy-current flowmetering method which is applicable to liquid metals.
The method works by exciting and then tracking eddy currents as they
are advected by a moving conductor. Because eddy currents decay by
about three orders of magnitude over the characteristic characteristic
magnetic diffusion time $\tau_{m}=\mu_{0}\sigma R^{2},$ which is
$\sim\unit[0.1]{s}$ in a liquid metal with a characteristic conductivity
$\sigma\sim\unit[10^{6}]{S/m}$ and size $R\sim\unit[0.1]{m,}$ the
time interval over which they can be tracked is limited to a few magnetic
diffusion time scales $\tau_{m}.$ The distance over which the eddy
current pattern can be tracked scales as $\sim\Rm.$ This means that
for small $\Rm,$ the pick-up coils have to be placed sufficiently
close to the excitation coils at the distance $\sim\Rm$.

We considered several alternative measurement schemes in which different
characteristic features of eddy current distributions are tracked.
The trackable features are zero crossing points or extrema of the
emf induced by the transient eddy currents. There are two kinds of
extrema which can be tracked: temporal and spacial. The former corresponds
to emf $(\mathcal{E})$ passing through extremum in time at a fixed
axial position. The latter corresponds to emf passing through extremum
at some axial position at a fixed instant of time. Mathematically,
these two types of extrema are defined as zero crossings of $\partial_{t}\mathcal{E}$
and $\partial_{z}\mathcal{E},$ respectively. Physically, temporal
extremum can be detected using a single pick-up coil, whereas at least
two coils are required to detect the passage of spatial extremum. 

A zero crossing point is detected by the the original measurement
scheme of \citet{Zheigur1965} using a single pick-up coil and its
distance from the excitation coil to determine the flow rate. The
location of excitation coil is not required by the measurement schemes
of \citet{Forbriger2015} and \citet{Krauter2017a} who use respectively
three and two non-coaxial pick-up coils to track approximately the
emf zero crossing point. The measurement scheme analysed numerically
by \citet{Tarabad1983} is based on the detection of temporal emf
extremum using a single excitation coil and two symmetric pick-up
coils in the differential connection.

In the mono-harmonic eddy current distributions, which were analyzed
first, the spatial extrema of emf move in the same way as zero crossings
because both remain separated by a quarter wave length. In this case,
the velocity of medium can be determined simply as $v=z/t,$ where
$t$ is the time after the eddy current excitation at which $\mathcal{E}$
or $\partial_{z}\mathcal{E}$ passes through zero at the distance
$z$ from the wave node or its extremum, respectively. Flowmetering
using temporal extrema of emf is slightly more complicated because
these extrema occur some time after zero crossing. This additional
time delay, which shows up also in the numerical results of \citet{Tarabad1983}
and significantly disturbs their measurement scheme, depends on the
conductivity of medium as well as on the eddy current distribution
but not on the position of the observation point provided that it
is not too close to the initial zero crossing point. Therefore the
additional time delay can be eliminated by using two pick-up coils
placed at $z_{1}$ and $z_{2}.$ Then the velocity of the medium can
be found as $v=(z_{2}-z_{1})/(t_{2}-t_{1}),$ where $t_{1}$ and $t_{2}$
are the times at which temporal extrema are detected in the respective
coil. The same approach can be used to eliminate the uncertainty in
the emf zero crossing time caused by the inaccuracy in the position
of excitation coils. Note that the measurement scheme of \citet{Krauter2017a}
is somewhat different because it relies on the assumption that the
emf varies linearly in the space between the two pick-up coils. \citet{Forbriger2015}
make a similar assumption about the spatial variation of the magnetic
flux density. These assumptions, which may hold for sufficiently closely
spaced pick-up coils but not in general, are not required in the measurement
schemes considered in this study.

We considered also more realistic eddy current distributions generated
by two anti-symmetric circular current loops or a single loop. In
the anti-symmetric set-up, the zero crossing point of emf as well
as the subsequent temporal extremum was found to travel synchronously
with the medium in the same way as with the mono-harmonic wave considered
before. But this was not the case for the two spatial extrema which
appear at both current loops in this set-up. These two extrema were
found to move at substantially different velocities from that of the
medium. This result highlights the crucial importance of symmetry
which holds for zero crossing points of emf but not for the two spatial
extrema in the anti-symmetric set-up. In a single-loop set-up, which
generates a spatially symmetric eddy current distribution, the spatial
extremum of emf was found to travel synchronously with the medium
as in the mono-harmonic wave. In this set-up, the velocity of the
medium can be determined by also tracking axial extremum of the magnetic
flux, which coincides with the zero crossing of the radial component
of the magnetic field. It has to be noted that because of the initial
tilt of the magnetic flux lines in the direction motion, the extremum
of magnetic flux arrives at a given observation point ahead that of
emf. This time lead can be eliminated similarly to the delay of temporal
extremum of emf by using two sensors as discussed above.

Finally, we analyzed the effect of a possible current asymmetry in
the two-loop set-up, and showed that it gives rise to a spurious drift
of the emf zero crossing point. It means that the mutual symmetry
of exiting coils  is crucial for the transient-eddy flowmetering.
Asymmetry of a few per cent was found to result in the zero drift
with a dimensionless velocity $\Rm\sim0.1.$ For the characteristic
parameters used at the beginning of this section, the respective physical
velocity is $v\sim\unit[0.1]{m/s.}$ It means that with this level
of asymmetry, which is not unlikely in practice, transient eddy current
flowmetering can be reliable only for the flows with $\Rm\gtrsim0.1.$
This estimate is consistent with the lowest $\Rm\approx0.35$ and
$\Rm\approx0.11$ achieved respectively by \citet{Zheigur1965} and
\citet{Forbriger2015}. At lower velocities, a more accurate symmetry
adjustment or calibration of the device may be required. This obviously
applies not only to the axisymmetric systems considered in our study
but also to more complex non-coaxial coil arrengements used in the
previous experimental studies.

The results of this study may be useful for designing more accurate
and reliable transient eddy-current flowmeters for liquid metals.

\section*{References}

\bibliographystyle{elsarticle-num-names}
\bibliography{tecfm}

\begin{thebibliography}{28}
\providecommand{\natexlab}[1]{#1}
\providecommand{\url}[1]{\texttt{#1}}
\providecommand{\urlprefix}{URL }
\expandafter\ifx\csname urlstyle\endcsname\relax
  \providecommand{\doi}[1]{doi:\discretionary{}{}{}#1}\else
  \providecommand{\doi}[1]{doi:\discretionary{}{}{}\begingroup
  \urlstyle{rm}\url{#1}\endgroup}\fi
\providecommand{\bibinfo}[2]{#2}

\bibitem[{Schulenberg and Stieglitz(2010)}]{Schulenberg2010}
\bibinfo{author}{T.~Schulenberg}, \bibinfo{author}{R.~Stieglitz},
  \bibinfo{title}{Flow measurement techniques in heavy liquid metals},
  \bibinfo{journal}{Nucl. Eng. Des.}
  \bibinfo{volume}{240}~(\bibinfo{number}{9}) (\bibinfo{year}{2010})
  \bibinfo{pages}{2077--2087}.

\bibitem[{Eckert et~al.(2011)Eckert, Buchenau, Gerbeth, Stefani, and
  Weiss}]{Eckert2011}
\bibinfo{author}{S.~Eckert}, \bibinfo{author}{D.~Buchenau},
  \bibinfo{author}{G.~Gerbeth}, \bibinfo{author}{F.~Stefani},
  \bibinfo{author}{F.-P. Weiss}, \bibinfo{title}{Some recent developments in
  the field of measuring techniques and instrumentation for liquid metal
  flows}, \bibinfo{journal}{J. Nucl. Sci. Technol.}
  \bibinfo{volume}{48}~(\bibinfo{number}{4}) (\bibinfo{year}{2011})
  \bibinfo{pages}{490--498}.

\bibitem[{Poornapushpakala et~al.(2014)Poornapushpakala, Gomathy, Sylvia, and
  Babu}]{Poornapushpakala2014}
\bibinfo{author}{S.~Poornapushpakala}, \bibinfo{author}{C.~Gomathy},
  \bibinfo{author}{J.~Sylvia}, \bibinfo{author}{B.~Babu},
  \bibinfo{title}{Design, development and performance testing of fast response
  electronics for eddy current flowmeter in monitoring sodium flow},
  \bibinfo{journal}{Flow Meas. Instrum.} \bibinfo{volume}{38}
  (\bibinfo{year}{2014}) \bibinfo{pages}{98--107}.

\bibitem[{Hussain and Baker(1985)}]{Hussain1985}
\bibinfo{author}{Y.~Hussain}, \bibinfo{author}{R.~Baker},
  \bibinfo{title}{Optimised noncontact electromagnetic flowmeter},
  \bibinfo{journal}{J. Phys. E: Sci. Instrum.}
  \bibinfo{volume}{18}~(\bibinfo{number}{3}) (\bibinfo{year}{1985})
  \bibinfo{pages}{210}.

\bibitem[{McHale et~al.(1985)McHale, Hussain, Sanderson, and Hemp}]{McHale1985}
\bibinfo{author}{E.~J. McHale}, \bibinfo{author}{Y.~Hussain},
  \bibinfo{author}{M.~Sanderson}, \bibinfo{author}{J.~Hemp},
  \bibinfo{title}{Capacitively-coupled magnetic flowmeter},
  \bibinfo{howpublished}{US Patent 4,513,624}, \bibinfo{year}{1985}.

\bibitem[{Shercliff(1987)}]{Shercliff1987}
\bibinfo{author}{J.~Shercliff}, \bibinfo{title}{The Theory of Electromagnetic
  Flow-Measurement}, Cambridge Science Classics, \bibinfo{publisher}{Cambridge
  University Press}, ISBN \bibinfo{isbn}{9780521335546}, \bibinfo{year}{1987}.

\bibitem[{Thess et~al.(2007)Thess, Votyakov, Knaepen, and Zikanov}]{Thess2007a}
\bibinfo{author}{A.~Thess}, \bibinfo{author}{E.~Votyakov},
  \bibinfo{author}{B.~Knaepen}, \bibinfo{author}{O.~Zikanov},
  \bibinfo{title}{Theory of the Lorentz force flowmeter}, \bibinfo{journal}{New
  J. Phys.} \bibinfo{volume}{9}~(\bibinfo{number}{8}) (\bibinfo{year}{2007})
  \bibinfo{pages}{299}.

\bibitem[{Wegfrass et~al.(2012)Wegfrass, Diethold, Werner, Fr{\"o}hlich,
  Halbedel, Hilbrunner, Resagk, and Thess}]{Wegfrass2012}
\bibinfo{author}{A.~Wegfrass}, \bibinfo{author}{C.~Diethold},
  \bibinfo{author}{M.~Werner}, \bibinfo{author}{T.~Fr{\"o}hlich},
  \bibinfo{author}{B.~Halbedel}, \bibinfo{author}{F.~Hilbrunner},
  \bibinfo{author}{C.~Resagk}, \bibinfo{author}{A.~Thess}, \bibinfo{title}{A
  universal noncontact flowmeter for liquids}, \bibinfo{journal}{Appl Phys
  Lett} \bibinfo{volume}{100}~(\bibinfo{number}{19}) (\bibinfo{year}{2012})
  \bibinfo{pages}{194103}.

\bibitem[{Priede et~al.(2009)Priede, Buchenau, and Gerbeth}]{Priede2009b}
\bibinfo{author}{J.~Priede}, \bibinfo{author}{D.~Buchenau},
  \bibinfo{author}{G.~Gerbeth}, \bibinfo{title}{Force-free and contactless
  sensor for electromagnetic flowrate measurements},
  \bibinfo{journal}{Magnetohydrodynamics}
  \bibinfo{volume}{45}~(\bibinfo{number}{3}) (\bibinfo{year}{2009})
  \bibinfo{pages}{451--458}.

\bibitem[{Shercliff(1960)}]{Shercliff1960}
\bibinfo{author}{J.~Shercliff}, \bibinfo{title}{Improvements in or relating to
  electromagnetic flowmeters}, \bibinfo{howpublished}{GB Patent 831226},
  \bibinfo{year}{1960}.

\bibitem[{Bucenieks(2005)}]{Bucenieks2005}
\bibinfo{author}{I.~Bucenieks}, \bibinfo{title}{Modelling of rotary inductive
  electromagnetic flowmeter for liquid metals flow control}, in:
  \bibinfo{booktitle}{Proc. 8th Int. Symp. on Magnetic Suspension Technology,
  Dresden, Germany, 26--28 September}, \bibinfo{pages}{204--8},
  \bibinfo{year}{2005}.

\bibitem[{Buchenau et~al.(2014)Buchenau, Galindo, and Eckert}]{Buchenau2014}
\bibinfo{author}{D.~Buchenau}, \bibinfo{author}{V.~Galindo},
  \bibinfo{author}{S.~Eckert}, \bibinfo{title}{The magnetic flywheel flow
  meter: Theoretical and experimental contributions}, \bibinfo{journal}{Appl.
  Phys. Lett.} \bibinfo{volume}{104}~(\bibinfo{number}{22})
  (\bibinfo{year}{2014}) \bibinfo{pages}{223504}.

\bibitem[{Hvasta et~al.(2018)Hvasta, Dudt, Fisher, and Kolemen}]{Hvasta2018}
\bibinfo{author}{M.~Hvasta}, \bibinfo{author}{D.~Dudt},
  \bibinfo{author}{A.~Fisher}, \bibinfo{author}{E.~Kolemen},
  \bibinfo{title}{Calibrationless rotating Lorentz-force flowmeters for low
  flow rate applications}, \bibinfo{journal}{Meas. Sci. Technol.}
  \bibinfo{volume}{29}~(\bibinfo{number}{7}) (\bibinfo{year}{2018})
  \bibinfo{pages}{075303}.

\bibitem[{Priede et~al.(2011{\natexlab{a}})Priede, Buchenau, and
  Gerbeth}]{Priede2011b}
\bibinfo{author}{J.~Priede}, \bibinfo{author}{D.~Buchenau},
  \bibinfo{author}{G.~Gerbeth}, \bibinfo{title}{Single-magnet rotary flowmeter
  for liquid metals}, \bibinfo{journal}{J. Appl. Phys.}
  \bibinfo{volume}{110}~(\bibinfo{number}{3})
  (\bibinfo{year}{2011}{\natexlab{a}}) \bibinfo{pages}{034512}.

\bibitem[{Lehde and Lang(1948)}]{Lehde1948}
\bibinfo{author}{H.~Lehde}, \bibinfo{author}{W.~Lang}, \bibinfo{title}{Device
  for measuring rate of fluid flow}, \bibinfo{howpublished}{US Patent
  2,435,043}, \bibinfo{year}{1948}.

\bibitem[{Cowley(1965)}]{Cowley1965}
\bibinfo{author}{M.~Cowley}, \bibinfo{title}{Flowmetering by a motion-induced
  magnetic field}, \bibinfo{journal}{J. Sci. Instrum.}
  \bibinfo{volume}{42}~(\bibinfo{number}{6}) (\bibinfo{year}{1965})
  \bibinfo{pages}{406}.

\bibitem[{Poornapushpakala et~al.(2010)Poornapushpakala, Gomathy, Sylvia,
  Krishnakumar, and Kalyanasundaram}]{Poornapushpakala2010a}
\bibinfo{author}{S.~Poornapushpakala}, \bibinfo{author}{C.~Gomathy},
  \bibinfo{author}{J.~Sylvia}, \bibinfo{author}{B.~Krishnakumar},
  \bibinfo{author}{P.~Kalyanasundaram}, \bibinfo{title}{An analysis on eddy
  current flowmeter--a review}, in: \bibinfo{booktitle}{Recent Advances in
  Space Technology Services and Climate Change (RSTSCC), 2010},
  \bibinfo{organization}{IEEE}, \bibinfo{pages}{185--188},
  \bibinfo{year}{2010}.

\bibitem[{Stefani et~al.(2004)Stefani, Gundrum, and Gerbeth}]{Stefani2004}
\bibinfo{author}{F.~Stefani}, \bibinfo{author}{T.~Gundrum},
  \bibinfo{author}{G.~Gerbeth}, \bibinfo{title}{Contactless inductive flow
  tomography}, \bibinfo{journal}{Phys. Rev. E}
  \bibinfo{volume}{70}~(\bibinfo{number}{5}) (\bibinfo{year}{2004})
  \bibinfo{pages}{056306}.

\bibitem[{Stefani and Gerbeth(2000)}]{Stefani2000}
\bibinfo{author}{F.~Stefani}, \bibinfo{author}{G.~Gerbeth}, \bibinfo{title}{A
  contactless method for velocity reconstruction in electrically conducting
  fluids}, \bibinfo{journal}{Meas. Sci. Technol.}
  \bibinfo{volume}{11}~(\bibinfo{number}{6}) (\bibinfo{year}{2000})
  \bibinfo{pages}{758}.

\bibitem[{Feng et~al.(1975)Feng, Deeds, and Dodd}]{Feng1975}
\bibinfo{author}{C.~Feng}, \bibinfo{author}{W.~Deeds},
  \bibinfo{author}{C.~Dodd}, \bibinfo{title}{Analysis of eddy-current
  flowmeters}, \bibinfo{journal}{J Appl Phys}
  \bibinfo{volume}{46}~(\bibinfo{number}{7}) (\bibinfo{year}{1975})
  \bibinfo{pages}{2935--2940}.

\bibitem[{Priede et~al.(2011{\natexlab{b}})Priede, Buchenau, and
  Gerbeth}]{Priede2011c}
\bibinfo{author}{J.~Priede}, \bibinfo{author}{D.~Buchenau},
  \bibinfo{author}{G.~Gerbeth}, \bibinfo{title}{Contactless electromagnetic
  phase-shift flowmeter for liquid metals}, \bibinfo{journal}{Meas. Sci.
  Technol.} \bibinfo{volume}{22}~(\bibinfo{number}{5})
  (\bibinfo{year}{2011}{\natexlab{b}}) \bibinfo{pages}{055402}.

\bibitem[{Vir{\'e} et~al.(2010)Vir{\'e}, Knaepen, and Thess}]{Vire2010}
\bibinfo{author}{A.~Vir{\'e}}, \bibinfo{author}{B.~Knaepen},
  \bibinfo{author}{A.~Thess}, \bibinfo{title}{Lorentz force velocimetry based
  on time-of-flight measurements}, \bibinfo{journal}{Phys Fluids}
  \bibinfo{volume}{22}~(\bibinfo{number}{12}) (\bibinfo{year}{2010})
  \bibinfo{pages}{125101}.

\bibitem[{Looney and Priede(2018)}]{Looney2017}
\bibinfo{author}{R.~Looney}, \bibinfo{author}{J.~Priede},
  \bibinfo{title}{Concept of a next-generation electromagnetic phase-shift
  flowmeter for liquid metals}, \bibinfo{journal}{(submitted to Flow Meas.
  Instrum.)} .

\bibitem[{Zheigur and Sermons(1965)}]{Zheigur1965}
\bibinfo{author}{B.~D. Zheigur}, \bibinfo{author}{G.~Y. Sermons},
  \bibinfo{title}{Pulse method of measuring the rate of flow of a conducting
  fluid}, \bibinfo{journal}{Magnetohydrodynamics}
  \bibinfo{volume}{1}~(\bibinfo{number}{1}) (\bibinfo{year}{1965})
  \bibinfo{pages}{101--104}.

\bibitem[{Tarabad and Baker(1983)}]{Tarabad1983}
\bibinfo{author}{M.~Tarabad}, \bibinfo{author}{R.~C. Baker},
  \bibinfo{title}{Computation of pulsed field electromagnetic flowmeter
  response to profile change}, \bibinfo{journal}{J. Phys. D: Appl. Phys.}
  \bibinfo{volume}{16}~(\bibinfo{number}{11}) (\bibinfo{year}{1983})
  \bibinfo{pages}{2103}.

\bibitem[{Forbriger and Stefani(2015)}]{Forbriger2015}
\bibinfo{author}{J.~Forbriger}, \bibinfo{author}{F.~Stefani},
  \bibinfo{title}{Transient eddy current flow metering},
  \bibinfo{journal}{Meas. Sci. Technol.}
  \bibinfo{volume}{26}~(\bibinfo{number}{10}) (\bibinfo{year}{2015})
  \bibinfo{pages}{105303}.

\bibitem[{Krauter and Stefani(2017)}]{Krauter2017a}
\bibinfo{author}{N.~Krauter}, \bibinfo{author}{F.~Stefani},
  \bibinfo{title}{Immersed transient eddy current flow metering: a
  calibration-free velocity measurement technique for liquid metals},
  \bibinfo{journal}{Meas. Sci. Technol.}
  \bibinfo{volume}{28}~(\bibinfo{number}{10}) (\bibinfo{year}{2017})
  \bibinfo{pages}{105301}.

\bibitem[{Abramowitz and Stegun(1972)}]{Abramowitz1964}
\bibinfo{author}{M.~Abramowitz}, \bibinfo{author}{I.~A. Stegun},
  \bibinfo{title}{Handbook of Mathematical Functions},
  \bibinfo{publisher}{Dover}, \bibinfo{address}{New York},
  \bibinfo{year}{1972}.

\end{thebibliography}

\end{document}